\pdfoutput=1
\documentclass[11pt]{article}

\usepackage[final]{acl}
\usepackage{verbatim}
\usepackage{times}
\usepackage{latexsym}
\usepackage{booktabs}
\usepackage[T1]{fontenc}
\usepackage[utf8]{inputenc}

\usepackage{microtype}

\usepackage{inconsolata}

\usepackage{graphicx}
\usepackage{amsmath} 
\usepackage{xcolor}
\usepackage{natbib}
\usepackage{pgffor}
\usepackage{graphicx}
\usepackage{placeins}    % for \FloatBarrier
\usepackage{float}       % if you might want [H]
\definecolor{darkgreen}{RGB}{0,100,0} 
\usepackage{threeparttable}

\AtBeginDocument{%
  }

\title{Pathway to Relevance:\\How Cross-Encoders Implement a Semantic Variant of BM25}

\author{
Meng Lu\thanks{Equal contribution.} \\
Brown University \\   
\texttt{meng\_lu@brown.edu}
\And
Catherine Chen\footnotemark[1] \\
Brown University \\
\texttt{catherine\_s\_chen@brown.edu}
\AND
Carsten Eickhoff \\
University of T\"ubingen \\
\texttt{carsten.eickhoff@uni-tuebingen.de}
}

\begin{document}
\maketitle
\begin{abstract}

Mechanistic interpretation has greatly contributed to a more detailed understanding of generative language models, enabling significant progress in identifying structures that implement key behaviors through interactions between internal components. In contrast, interpretability in information retrieval (IR) remains relatively coarse-grained, and much is still unknown as to how IR models determine whether a document is relevant to a query. In this work, we address this gap by mechanistically analyzing how one commonly used model, a cross-encoder, estimates relevance. We find that the model extracts traditional relevance signals, such as term frequency and inverse document frequency, in early-to-middle layers. These concepts are then combined in later layers, similar to the well-known probabilistic ranking function, BM25. Overall, our analysis offers a more nuanced understanding of how IR models compute relevance. Isolating these components lays the groundwork for future interventions that could enhance transparency, mitigate safety risks, and improve scalability.
\end{abstract}

\section{Introduction}

Information retrieval (IR) is the subfield of NLP that aims to rank a collection of documents by their relevance to a specific query. Traditional ranking strategies have long relied on probabilistic models grounded in intuitive heuristics, like BM25 \citep{bm25}, to estimate relevance. BM25 leverages term frequency (TF) and inverse document frequency (IDF) to rank documents effectively, achieving strong performance across various tasks. Its simplicity and inherent interpretability have made it a cornerstone of traditional IR systems. Inspired by BM25’s success, earlier neural IR models \citep{pang2016text, guo2016deep} were purposefully designed to emulate BM25’s principles. These models incorporate explicit components for semantic TF and IDF computations, blending neural architectures with established IR heuristics to improve relevance estimation.

However, the advent of transformer-based models revolutionized the field of IR. These models, trained end-to-end on large numbers of query-document pairs~\citep{msmarco_dataset,beir}, excel at extracting context-dependent semantic signals for ranking tasks. By leveraging multi-headed attention and vast parameter spaces, transformers~\cite{vaswani_attention} capture nuanced relationships between query and document terms that go beyond traditional heuristic-based approaches. Despite their superior performance, these models come with significant trade-offs: their complexity and lack of interpretability make it challenging to understand their internal mechanisms. This raises fundamental questions: how do these models assess relevance? Do they rely on established IR principles such as TF and IDF, or do they draw on entirely different expressions of relevance?

In this work, we build upon previous correlational studies \citep{choi_idf, zhan, formal2021white, formal2022match, macavaney2022abnirml} by employing mechanistic interpretability methods~\citep{nanda2022mechanistic,elhage2021mathematical,olsson_induction,meng_rome,wang_ioi,pearl} to address these questions.

Concretely, we analyze a BERT-based IR model (i.e., a cross-encoder) to isolate multiple relevance signals beyond exact match TF.\footnote{All code and resources are available at: \url{https://github.com/mlu108/CrossEncoderBM25}}
Our main findings are:
(1) We identify attention heads that extract and process BM25-like components, including a semantic version of term frequency (soft-TF), term saturation, and document length (\S\ref{sec:relevance_scoring_heads}-\S\ref{matching_heads}). 
(2) We find evidence that IDF information largely already exists in the largest singular value of the embedding matrix and can be manipulated to later control term importance (\S\ref{IDF}).
(3) We confirm that heads in later layers aggregate all these relevance signals in a BM25-style manner by defining a linear approximation of the hypothesized relevance computation and evaluating its ability to reconstruct cross-encoder's predicted scores, confirming that our circuit captures the core mechanism of relevance computation (\S\ref{sec:circuit-function-approximatio}).

Overall, our mechanistic analysis uncovers insights on how IR models determine relevance, paving the way for targeted model editing to boost performance, enable personalization, and mitigate bias in transformer‑based IR.

\section{Background and Related Work} \label{background}

\subsection{Axioms and BM25}\label{axioms_bm25}

\begin{table*}[h!]
  \small
  \centering
  \begin{tabular}{p{1.6cm} p{3.8cm} p{5cm} p{4cm}}
    \toprule
    
    \textbf{BM25\newline Component} & \textbf{Axiom} & \textbf{Perturbation} & \textbf{Axiom-Based Diagnostic Dataset Examples }\newline(\textit{b}: baseline doc, \textit{p}: perturbed doc) \\
    \midrule
    \textbf{TF} (\textit{soft-TF}) 
    & \textbf{TFC1~\cite{fang2004formal}}: Prefer documents with more query term occurrences. 
    & Given a baseline document, we perturb it by appending one more selected query term. 
    & \textbf{\textit{b}:} Quebec is a small city in Canada. \newline 
      \textbf{\textit{p}:} Quebec is a small city in Canada. \textcolor{darkgreen}{Quebec}. \\
    \cmidrule(l){2-4}
    & \textbf{STMC1~\cite{fang2006semantic}}: Prefer documents with semantically similar terms. 
    & Given a baseline document, we append a semantically similar term - the synonym with highest embedding cosine similarity with query term from 20 candidates generated by GPT-4o.
    & \textbf{\textit{b}:} Quebec is a small city in Canada. \newline 
      \textbf{\textit{p}:} Quebec is a small city in Canada. \textcolor{darkgreen}{Toronto}. \\
    \midrule
    \textbf{IDF} 
    & \textbf{TDC~\cite{fang2004formal}}: Prefer documents with more discriminative query terms. 
    & N/A  
    & N/A \newline 
      \\
    \midrule
    \textbf{k} 
    & \textbf{TFC2~\cite{fang2004formal}}: Additional occurrences yield smaller improvements. 
    & We create a baseline document using GPT-4o to create five relevant sentences starting with selected query term's pronoun and perturb by incrementally restoring the term.
    & \textbf{\textit{b}:} It is in Canada. It is fun. \newline 
      \textbf{\textit{p\textsubscript{1}}:} \textcolor{darkgreen}{Quebec} is in Canada. It is fun. \newline 
      \textbf{\textit{p\textsubscript{2}}:} \textcolor{darkgreen}{Quebec} is in Canada. \textcolor{darkgreen}{Quebec} is fun. \\
    \midrule
    \textbf{b} 
    & \textbf{LNC1~\cite{fang2004formal}}: Penalize longer documents for non-relevant terms. 
    & Given a baseline document, we create five perturbations by sequentially appending an increasing number of random sentences from a non-relevant query in the base dataset.
    & \textbf{\textit{b}:} Quebec is in Canada. \newline 
      \textbf{\textit{p\textsubscript{1}}:} Quebec is in Canada. \textcolor{darkgreen}{Road not taken is written by Frost.} \newline 
      \textbf{\textit{p\textsubscript{2}}:} Quebec is in Canada. \textcolor{darkgreen}{Road not taken is written by Frost. Happiness is a Butterfly.} \\
    \bottomrule

  \end{tabular}
    \caption{Mapping of BM25 components to IR axioms. Note: Since IDF is a distinct component in BM25 (unlike term saturation and document length, which are tied to TF), we use alternative methods to isolate it (\S\ref{IDF}).}
\label{tab:axioms_table}
\end{table*}

Axiomatic IR constructs formal desiderata, or axioms, outlining specific properties that an effective ranking model should satisfy \citep{axioms_bruza}. For example, the TFC1 axiom \citep{fang2004formal} states that documents that contain more query terms should be ranked higher, and the TDC axiom \citep{fang2004formal} states that documents containing query terms with higher inverse document frequency (IDF) should receive higher scores. These axioms provide a theoretical foundation for understanding and developing ranking functions by formalizing human-interpretable and intuitive notions of relevance into mathematical constraints. 

BM25 is a widely used probabilistic ranking function that exemplifies these axiomatic principles by ranking documents based on term frequency (TF), inverse document frequency (IDF), term saturation, and document length. It is defined as:

\begin{equation}
 \sum_{t \in q} \text{IDF}(t) \cdot \frac{\text{TF}(t,d) \cdot (k_1 + 1)}{\text{TF}(t,d) + k_1 \cdot (1 - b + b \cdot \frac{|d|}{\text{avgdl}})}
\end{equation}

where $q$ is the query, $d$ is the document, $t$ is a query term, $k_1$ and $b$ are hyperparameters controlling term saturation and length normalization, respectively. $|d|$ is the length of document $d$, and $\text{avgdl}$ is the average document length in the corpus. 

Intuitively, BM25 computes a relevance score by multiplying each query term’s term frequency (TF) in the document with its inverse document frequency (IDF), capturing both how relevant and how informative the term is. It then sums these scores across all query terms with two adjustments: diminishing returns for repeated terms (term saturation via $k_1$), and downweighting for longer documents (length normalization via $|d|$), which prevents overly long documents with additional query term occurrences from being unfairly favored. As shown in Table \ref{tab:axioms_table}, each component of BM25 aligns with a specific axiom. For instance, TFC1 reflects the TF factor, which favors documents containing more query terms. Framing BM25 in terms of axioms allows us to mechanistically analyze whether and how neural IR models internalize these principles when estimating relevance.

\subsection{Mechanistic Interpretability for IR}

Despite not being explicitly trained to encode traditional relevance concepts, previous work provides correlational evidence suggesting that BERT-based IR models encode BM25-like information ~\citep{wang2021bert, rau2022different, yates2021pretrained, macavaney2022abnirml, choi_idf, zhan, formal2021white, formal2022match}. 

To move beyond correlation and gain a causal understanding of how relevance is computed, we turn to mechanistic interpretability~\citep{nanda2022mechanistic, elhage2021mathematical, pearl}, which has been instrumental in uncovering how transformer-based NLP models perform certain tasks, such as indirect object identification \cite{meng_rome} and greater-than computation \cite{hanna2024does}. 

In the context of IR, \citet{chen_axiomatic} apply activation patching~\citep{Vig, meng_rome} to identify attention heads responsible for exact term matching. We build on this work by decomposing BM25 into intuitive IR axioms and constructing diagnostic datasets to localize each component. Rather than focusing solely on exact matches, we extend the analysis to semantic matches and trace how these components interact to compute the final relevance score using path patching~\citep{wang_ioi, goldowskydill_ppatching}. This work is the first to causally uncover an internal circuit for the relevance estimation mechanism of neural IR architectures.

\section{Methodology}\label{sec:method}
\subsection{Model}

While many early neural models were purposefully designed to emulate BM25's principles, transformer-based models have revolutionized the field of IR. Among them, cross-encoders represent a specific architectural approach to neural ranking models that process query-document pairs jointly.

Given input \texttt{x} in the format: \texttt{<CLS> query <SEP> document <SEP>}, the model is trained to optimize a binary classification task where the \texttt{<CLS>} token is passed to a classifier to determine whether the provided query is relevant to the document.
In this work, we choose to examine ms-marco-MiniLM-L-12-v2~\citep{minilm_huggingface}, a BERT-based cross-encoder, for its high performance on common IR benchmarks~\citep{msmarco_dataset,trec_dataset}.

\subsection{Diagnostic Datasets Construction} 

To investigate if and how the model implements BM25, we map BM25 components to IR axioms (§\ref{axioms_bm25}) and construct diagnostic datasets that isolate individual axiomatic components. We build off of~\citet{chen_axiomatic}, who create a diagnostic dataset for analyzing TFC1 using MS-MARCO \citep{msmarco_dataset}, consisting of 10k web search query-document pairs. To create a diagnostic dataset for the remaining axioms, we perturb each query-document pair in this base dataset following their formal axiomatic definition (Table \ref{tab:axioms_table}). Each perturbed sample differs minimally from the original, with the perturbation introducing an additional signal corresponding to the axiom. Our perturbation strategies and examples of input pairs are shown in Table \ref{tab:axioms_table}. Additionally, we expand the analysis on TF to \textit{soft-TF} to include semantic matches, motivated by findings that neural models often score documents highly even in the absence of exact lexical matches~\citep{rau2022different}.

\subsection{Path Patching on Diagnostic Datasets} 

We first begin by tracking term matching, the core feature of classical IR models like BM25, by applying path patching on minimal pairs in TFC1 and STMC1 diagnostic datasets to uncover the soft-TF mechanism. For other components in BM25, we use these datasets to further inspect how they interact with the main soft-TF circuit.

Specifically, we isolate soft-TF by selecting a query, a baseline document \textit{b}, and a perturbed document \textit{p}, where \textit{p} introduces an additional semantic or lexical term frequency signal. We then substitute activations from \textit{p} into \textit{b} at specific model components, such as individual attention heads or MLP layers. If this substitution causes the model’s relevance score for \textit{b} to shift toward that of \textit{p}, we interpret it as causal evidence that the patched component encodes a term matching signal.

The path patching algorithm proceeds iteratively in a backward fashion: it first identifies components directly responsible for changes in the output logits, then recursively traces upstream to uncover the full set of causally relevant components. This backward search reveals the components through which specific information, introduced by the perturbation, flows (more details in Appendix \ref{sec:path-patching-method}).

\section{Semantic Scoring Circuit} \label{sec:circuit_section}

\begin{figure}
    \centering
    \includegraphics[width=0.9\columnwidth]{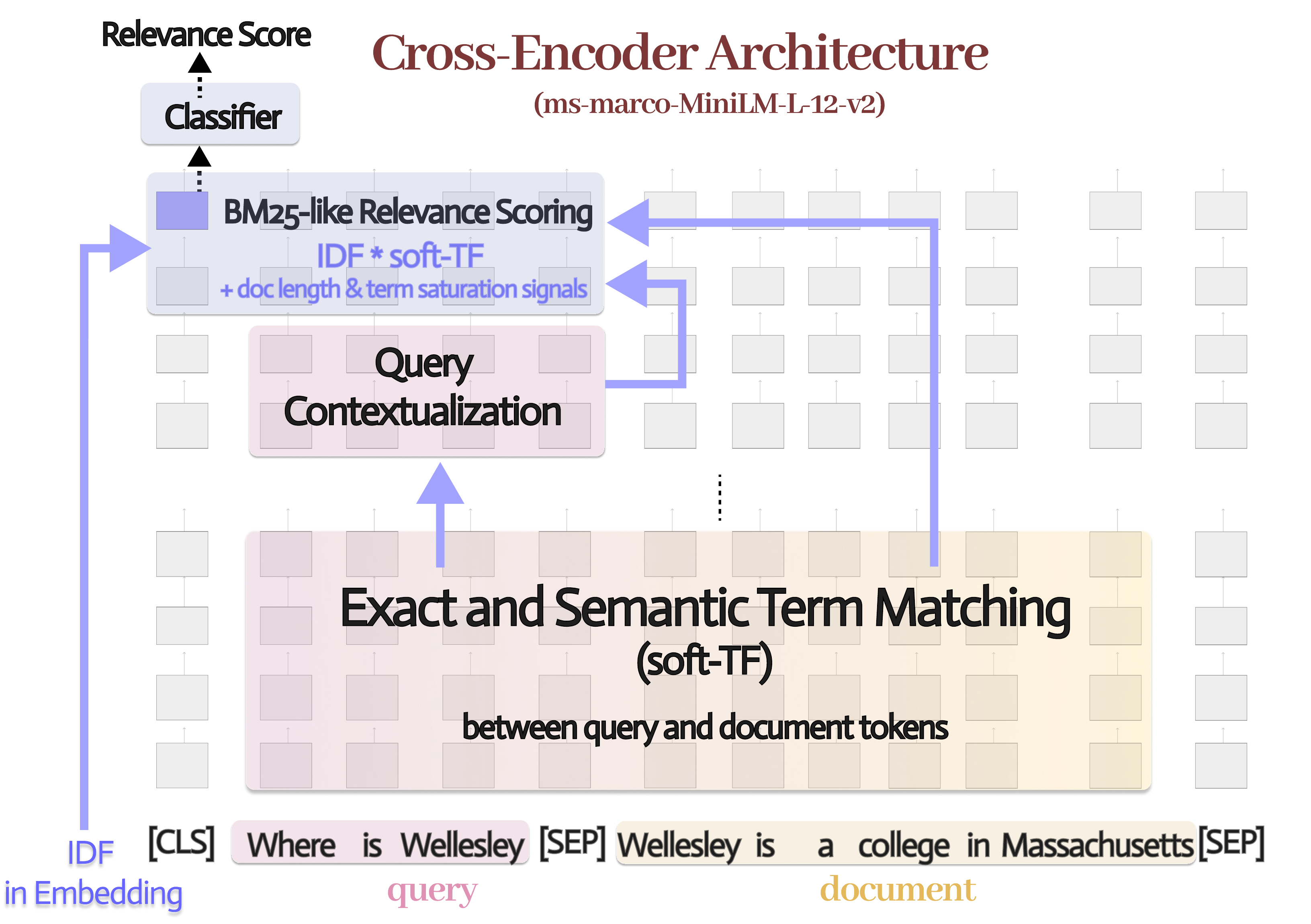}
    \caption{Overview of relevance mechanisms in the model. The model first jointly analyzes query and document tokens to identify matching terms (exact and semantic), then contextualizes each query term within the full query, and finally calculates a relevance score by weighting query terms by their importance (IDF) and aggregating them, similar to BM25.}

    \label{fig:main_figure}
    \vspace{-0.5em}
\end{figure}

In the course of this section, we will identify and localize the following components in the model's \textit{Semantic Scoring Circuit} (overview in Figure \ref{fig:main_figure}, detailed walk-through example in Figure \ref{fig:pipeline-example-overview}):

\begin{itemize}
    \item \textit{Matching Heads} locate exact and semantic term matches in the document for each query term, while also encoding term saturation and document length signals.
    \item \textit{Query Contextualization Heads} distribute the matching signal from higher-IDF query tokens to all query tokens.
    \item \textit{IDF} of each term is stored in a dominant low-rank vector of the model's embedding matrix.
    \item \textit{Relevance Scoring Heads} aggregate query-term importance by combining the matching signal (soft-TF) and IDF for each query term in a manner similar to BM25.
\end{itemize}

As described in \S\ref{sec:method}, we begin by uncovering the core soft-TF mechanism via a ``backward pass'' of the model: path patching to the output logits, then progressively patching to earlier components to sequentially uncover important heads that compute and propagate soft-TF signals.

After localizing the soft-TF mechanism, we apply Singular Value Decomposition (SVD) on the embeddings to investigate IDF. After isolating all BM25-like components in this section, we verify that Relevance Scoring Heads perform a BM25-like computation in \S\ref{sec:circuit-function-approximatio}.

\subsection{Relevance Scoring Heads}
\label{sec:relevance_scoring_heads}
\begin{figure}[!ht]
    \centering
    \includegraphics[width=0.37\textwidth]{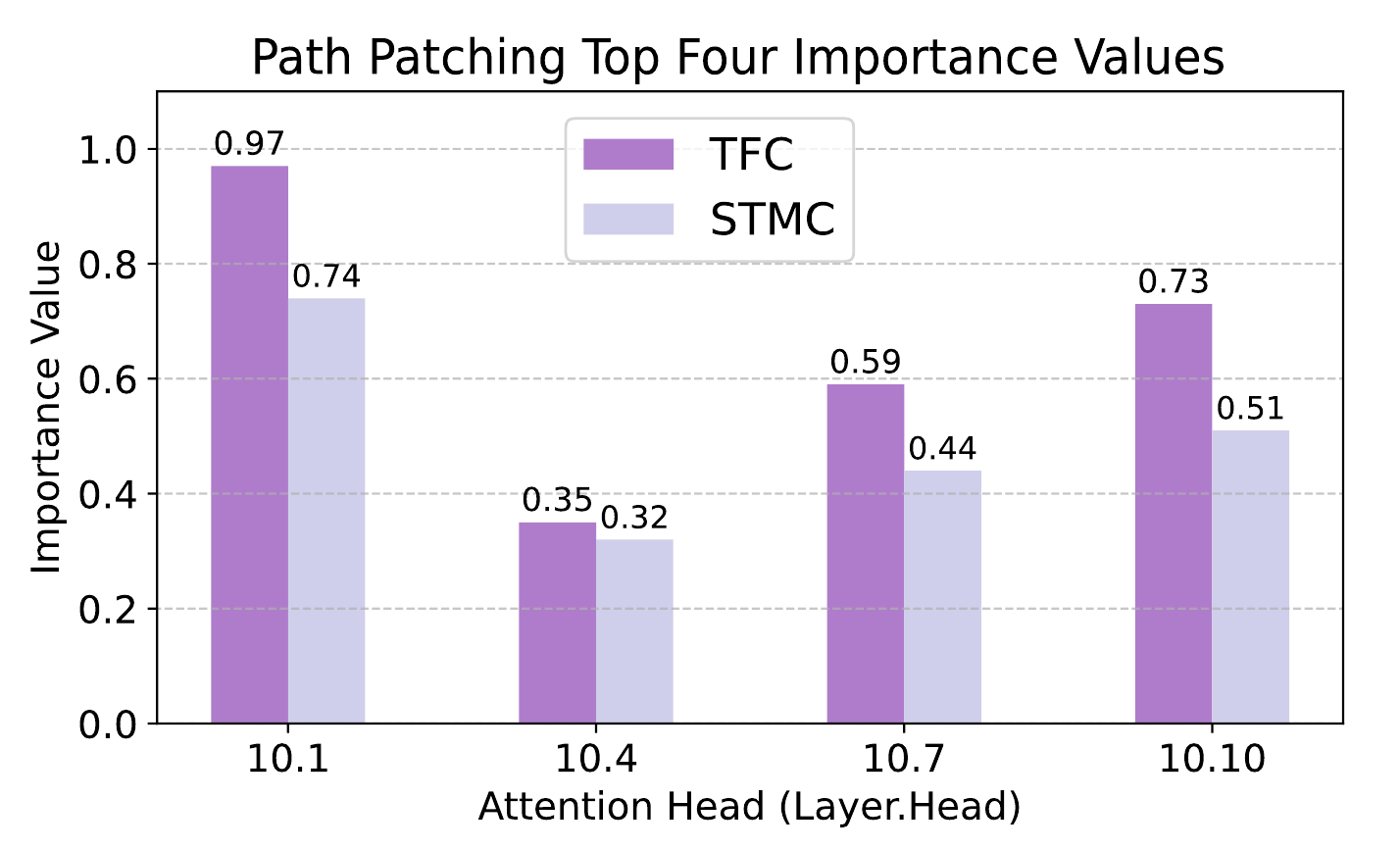}
    \caption{Path patching identifies heads 10.1, 10.4, 10.7, and 10.10 as the most important carriers of soft-TF signals to \texttt{[CLS]} on both TFC1 and STMC1, with similar patching effects.} 
    \label{fig:BMHeads_patching}
    % \vspace{-1em}
\end{figure}
%{paperdiagrams/patching_to_logits_barplot.png}
%\begin{figure}[!ht]
%    \centering
%    \includegraphics[width=0.35\textwidth]{paperdiagrams/path_patching_final_logits.pdf}
%    \caption{Path patching identifies heads 10.1, 10.4, 10.7, and 10.10 as key contributors of soft-TF signals to the \texttt{[CLS]} token for relevance scoring in both TFC1 and STMC1, with nearly identical patching effects across datasets.}
%    \footnotetext{We show heads in later layers because all earlier layers' heads have importance value below 0.005.}
%    \label{fig:BMHeads_patching}
    % \vspace{-1em}
%\end{figure}

\textbf{Information Flow.}
To identify the components that directly transmit soft-TF signals to the relevance scores, we path patch to the logits using the TFC1 and STMC1 diagnostic datasets. We observe a high Pearson correlation between the patching effects of TFC1 and STMC1 (\textit{r} = 0.99, \textit{p} < 0.001), indicating that path patching on both datasets identifies essentially the same set of influential heads. This suggests that the model treats exact and semantic matches similarly in the final relevance computation (see Figure~\ref{fig:appendix_path_patching_to_final_logits} in Appendix \ref{sec:appendix_detailed_ppatch} for detailed patching results). 

Figure~\ref{fig:BMHeads_patching} displays the heads with the highest patching effects that exceed the top 30\% of causal importance for both TFC1 and STMC1: 10.1 (Layer 10, Head 1), 10.4, 10.7, and 10.10. Notably, these heads contribute a larger change in relevance scores for TFC1 compared to STMC1, suggesting that the model may prioritize exact matches over semantic ones.

\begin{figure}[!ht]
    \centering
    \includegraphics[width=0.9\columnwidth]{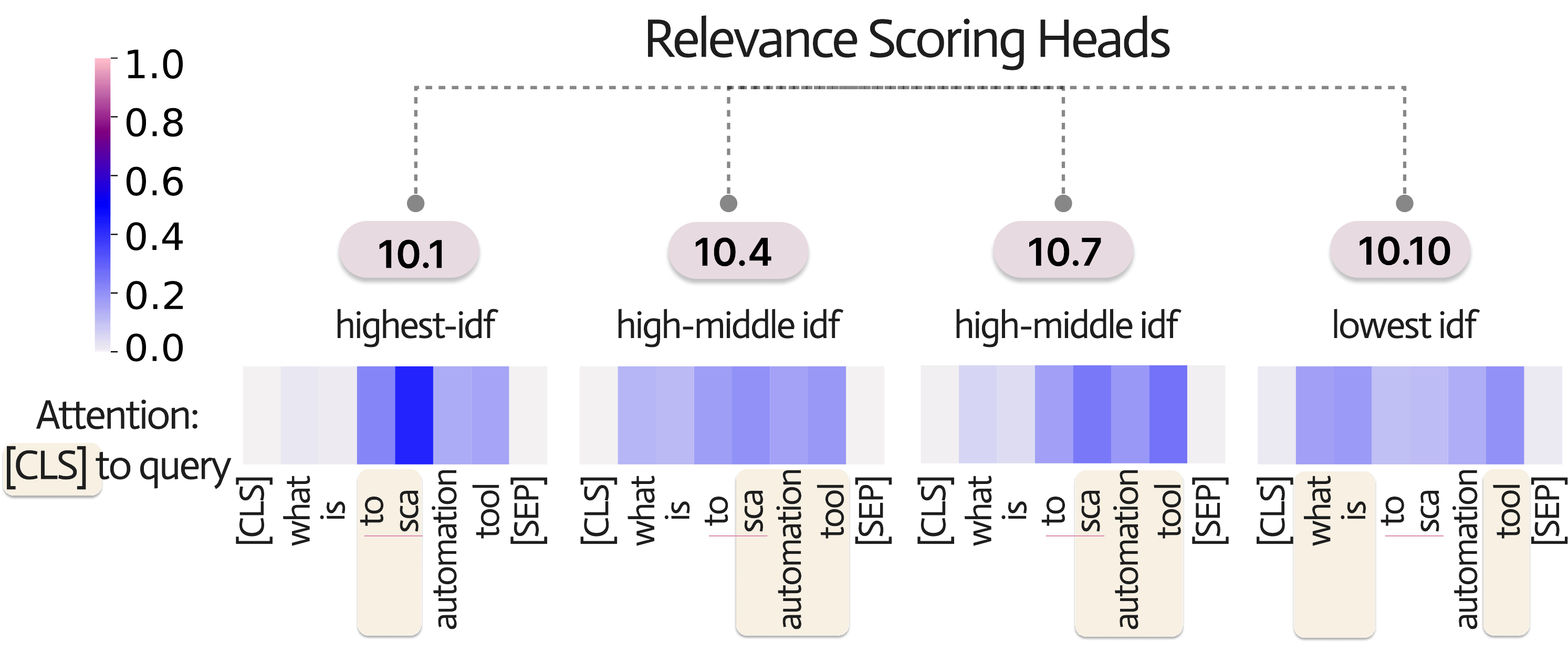}
    \caption{Example of the attention pattern from \texttt{[CLS]} to query tokens in the Relevance Scoring Heads, illustrating how these heads process \textit{soft-TF} for specific query tokens based on their IDF values.}

    \label{fig:BMHeads_IDF}
    % \vspace{-0.5em}
\end{figure}

\textbf{Component Behavior.}
Qualitative analysis reveals that the \texttt{[CLS]} token selectively attends to query tokens, indicating the movement of soft-TF signals into its representation to prepare for the final relevance score in the model's classification head. We find that the heads distribute attention to different parts of the query. Specifically, 10.1 focuses on high-IDF query terms, 10.10 on low-IDF, and 10.7 and 10.4 on mid-to-high-IDF (Figure~\ref{fig:BMHeads_IDF}). We verify this distributive behavior by calculating the Pearson correlation between each head's attention distribution and the IDF values of query tokens and find a moderately high average (\textit{r}=0.67, \textit{p} < 0.01).

\textbf{Comparison to BM25.}
Because these heads prepare the \texttt{[CLS]} token for the final scoring in the classification head, we call them \textit{Relevance Scoring Heads}. They combine IDF and soft-TF signals per query term, resembling BM25's term weighting mechanism. Thus, we hypothesize that the model combines the outputs of the Relevance Scoring Heads in a similar manner and verify this hypothesis in \S\ref{sec:circuit-function-approximatio}.

\textbf{Important Upstream Components.} 
We path patch to the Relevance Scoring Heads' value vectors to identify which upstream components transmit soft-TF signals. On both TFC1 and STMC1, patching effects show high average correlation of 0.83 (\textit{p} <0.001), revealing a shared set of attention heads, divided into two groups: (1) Query Contextualization Heads (\S\ref{sec:cqr_heads}) which redistribute soft-TF among query tokens and (2) Matching Heads (\S\ref{matching_heads}) which detect exact and semantic query-document matches (soft-TF) (see Appendix~\ref{sec:appendix_detailed_ppatch}).

\subsection{Query Contextualization Heads}
\label{sec:cqr_heads}

\textit{Query Contextualization Heads} (8.10 and 9.11) aggregate soft-TF signals of higher-IDF query tokens and distribute them across all query tokens, strengthening their representations for the Relevance Scoring Heads' final computation.

\textbf{Component Behavior.}
At the Relevance Scoring Heads, the \texttt{[CLS]} token retrieves soft-TF from all query tokens. Thus, we analyze the attention patterns among query tokens in these two intermediary heads to understand how they modify the query token representation in the residual stream. We find that all query tokens in these heads consistently focus on one or two higher-IDF query tokens, with strong correlations to IDF values (9.11: $r=0.829$, 8.10: $r=0.781$ ) with both \( p < 0.001 \). This suggests that 8.10 and 9.11 redistribute the soft-TF signals of higher-IDF tokens to contextualize the entire query for the Relevance Scoring Heads.

\textbf{Comparison to BM25.} 
BM25 weights each query term independently, with no exchange of information between terms. In contrast, Query Contextualization Heads seem to learn to contextualize the entire query, redistributing soft-TF signals to amplify high-IDF tokens and dynamically reweight terms, before the final scoring. Further study of this learned contextualization is left for future work.

\textbf{Important Upstream Components.} Path patching to 8.10 and 9.11 confirms that these heads receive soft-TF signals from the Matching Heads (more details in Appendix~\ref{sec:appendix_detailed_ppatch}).

\subsection{Matching Heads} \label{matching_heads}
\begin{figure}[!ht]
    \centering
    \includegraphics[width=0.9\columnwidth]{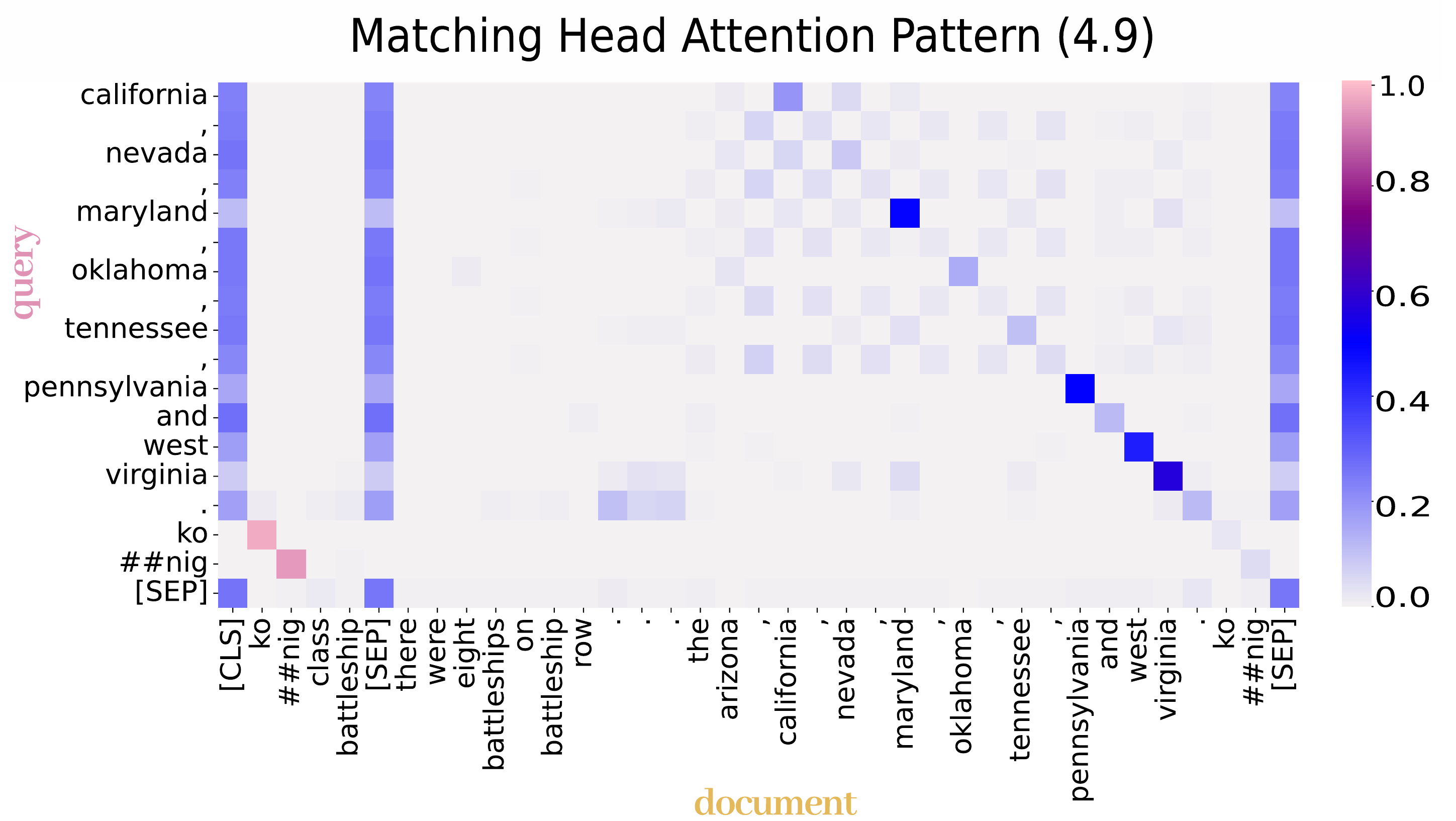}
    \caption{Example attention pattern of Matching Head 4.9: tokens attend most strongly to duplicates but also mildly to similar tokens.}
    \label{fig:MH_example}
    \vspace{-0.75em}
\end{figure}

Alongside Query Contextualization Heads, a separate set of heads, named \textit{Matching Heads} (0.8, 1.7, 2.1, 3.1, 4.9, 5.7, 5.9, 6.3, 6.5, 7.9, 8.0, 8.1, 8.8), in the model's early and middle layers detect exact and semantic token matches (soft-TF) between the query and document and pass these signals to the Relevance Scoring Heads.

\textbf{Component Behavior.} 
Qualitative inspection reveals that Matching Heads actively attend to both exact and semantically similar matches across the entire input sequence (i.e., query and document). As shown in Figure~\ref{fig:MH_example}, head 1.7 exhibits this behavior clearly: both query and document tokens strongly attend to duplicated terms and, to a lesser extent, to semantically related terms.

To quantitatively verify this behavior, we compute the Pearson correlation between attention weights and semantic similarity for each query–document token pair. If a head indeed matches semantically similar words, its attention from a query token $q_i$ to a document token $d_j$ should increase with $\cos(\mathbf{q}_i, \mathbf{d}_j)$. On average, Matching Heads exhibit a substantially stronger correlation ($r = 0.500$, $p < 0.001$) compared to other heads ($r = 0.132$). This evidence suggests that Matching Heads employ an attention mechanism that scales proportionally with semantic proximity: higher semantic similarity corresponds to higher attention scores. Since each attention weight quantifies the strength of token matching, the sum of attention values from a query token to all document tokens effectively approximates the soft-TF of a query term.

We validate the Matching Heads' importance by mean-ablating them, and find that the average logits across both TFC1 and STMC1 perturbed samples decrease (TFC1: 5.146 to -4.394, STMC1: -1.998 to -4.611), indicating that these heads are critical for computing a semantic version of term matches.

\textbf{Additional Relevance Signals.}
Since BM25's TF term takes two additional signals, term saturation and document length, into account, we also investigate whether they affect Matching Heads. To isolate these effects, we use two diagnostic datasets: (1) TFC2, which increases repeated query term occurrences to test for term saturation, and (2) LNC1, which adds irrelevant sentences to simulate the isolated effect of increased document length (Table~\ref{tab:axioms_table}).

Because each dataset varies only in one factor, we track each Matching Head’s average attention across these controlled groups. Consistent with how BM25 controls for term saturation and document length effects, we observe two trends (additional details in  Appendix~\ref{sec:matching_heads_additional_signals}): (1) in TFC2, attention to a term increases sharply after the initial occurrence, but plateaus with additional term repetitions, consistent with term saturation, and (2) in LNC1, attention decreases as irrelevant content increases, despite constant relevance, indicating sensitivity to document length. 

These findings suggest that Matching Heads integrate soft-TF with saturation and document-length signals rather than operating in isolation. We define this composite signal as the \textit{Matching Score}, reflecting its incorporation of soft term frequency, term saturation, and document length effects, three core signals of BM25.

\textbf{Comparison to BM25.} 
Matching Heads implement a semantic variant of the TF component in BM25 by identifying and weighting query-document token matches. Our TFC2 and LNC1 experiments show that they also go beyond simple match counting: their attention outputs are modulated in ways that mirror BM25’s additional relevance adjustments for term saturation and document length normalization. Thus, Matching Heads jointly capture all three core components of BM25, TF, term saturation, and length normalization, while also incorporating semantic similarity.

\textbf{Important Upstream Components.}
Path patching reveals that the embeddings are the primary input to Matching Heads (see Appendix~\ref{sec:ppatch_matching_heads}), confirming they generate soft-TF based on semantic similarity from the embeddings. This completes the backward tracing for the soft-TF circuit.

\subsection{IDF in the Embedding Matrix}
\label{IDF}

In addition to TF, IDF plays a critical role in BM25 by weighting each query token's TF contribution, allowing the model to prioritize more uncommon terms. As shown in \S\ref{sec:relevance_scoring_heads}, IDF similarly emerges as a key signal for the Relevance Scoring Heads in our model. While \citet{choi_idf} provide correlational evidence for IDF in the embedding matrix, we use SVD and low-rank interventions to causally localize their functional role in the model. 

SVD allows us to decompose a matrix into a sum of orthogonal rank-1 components, outer products of a column and row vector, ordered by their contribution to the overall matrix. Recall that the embedding matrix \( W_E \) is structured such that each row corresponds to a word's representation in the vocabulary, while the columns represent latent feature dimensions. By applying SVD to \( W_E \), we can analyze the dominant directions to see what signals (e.g., IDF) drive the model’s behavior. Mathematically, the SVD of \( W_E \) is expressed as:
\[
W_E = USV^T = \sum_{i=1}^r \sigma_i \, u_i \, v_i^T
\]

Here, \( \sigma_i \) are the singular values, which quantify the importance or strength of each rank-1 component; \( u_i \) and \( v_i \)  left and right singular vectors, respectively, forming orthonormal bases that capture patterns in the row (token embeddings) and column spaces (feature dimensions); and \( r \) is the rank of \( W_E \), representing the number of non-zero singular values and independent directions in the matrix. By focusing on the largest singular values (\( \sigma_i \)) and their corresponding singular vectors (\( u_i, v_i \)), we can study whether IDF is stored in dominant components of the embedding matrix. 

We find that the top singular vector \( U_0 \) is highly correlated (\textit{r} = \(-71.36\%\)) with MS-MARCO IDF values~\cite{msmarco_dataset}, the model's training dataset, indicating that IDF is encoded in the embedding matrix's dominant low-rank component.

\begin{figure}[!ht]
    \centering
    \includegraphics[width=0.35\textwidth]{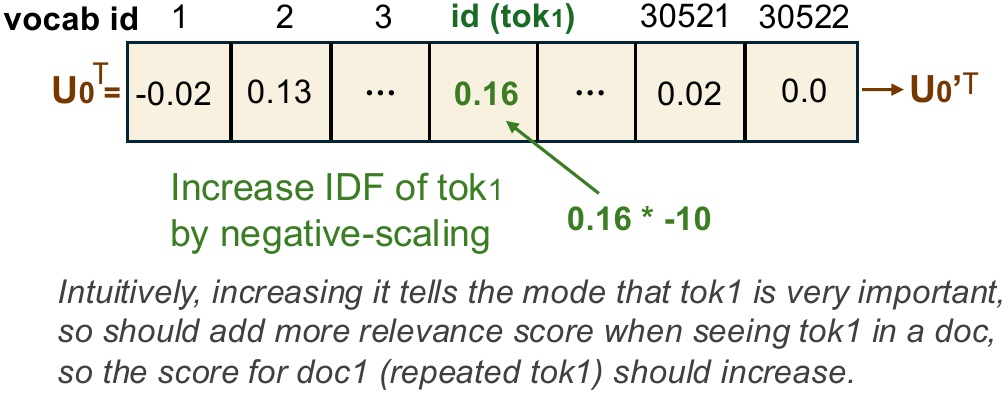}
    \caption{\(U_0\) editing example (transposed for visualization). Since \(U_0\) is negatively correlated with IDF, increasing \texttt{tok1}'s IDF, representing its importance in relevance computation, requires negatively scaling its \(U_0\) component.}
    \label{fig:model_edit_method}
    \vspace{-0.7em}
\end{figure}

\textbf{Causal Experiment.} 
To validate this correlation, we perform an intervention to demonstrate that the IDF values from \( U_0 \) have a causal effect on downstream components (i.e., relevance scoring heads) and thus, the overall relevance computation.

Given previous understanding on \( U_0 \), we can interpret \( U_0 \) as a 1-D IDF dictionary, where \( \text{idx}(q_i) \) corresponds to the vocabulary index of \( q_i \), mapping to its respective IDF value. Modifying the value at \( \text{idx}(q_i) \) in \( U_0 \) allows us to adjust the importance of the Matching Score for \( q_i \) (Figure~\ref{fig:model_edit_method}). 

If the model uses the IDF values encoded in \( U_0 \), then modifying these values should result in corresponding changes to the ranking score. Specifically, according to the BM25-based Scoring Hypothesis in \S\ref{sec:relevance_scoring_heads}, there is a linear relationship between the IDF of a query token \( q_i \) and the relevance score: increasing the IDF of \( q_i \) should increase the relevance score. We have shown that \( U_0 \) is rank-1 and negatively correlated with IDF. If the model indeed uses the IDF values encoded in \( U_0 \), we can increase \( \text{IDF}(q_i) \) by decreasing \( q_i \)'s value in \( U_0 \), which would directly increase the relevance score. The converse should hold true for decreasing IDF. 

To test this, we design an experiment with controlled, minimal examples. Given each query from our base dataset (e.g., ``computer science department number''), we create two documents where \texttt{doc1} repeats the first query token (e.g., ``computer computer computer'' and \texttt{doc2} repeats the second query token (e.g., ``science science science''). Then, we can edit the IDF of \texttt{tok1} and measure the effect on the relevance scores of \texttt{doc1} and \texttt{doc2}.

\begin{figure}[!ht]
    \centering
    \includegraphics[width=0.35\textwidth]{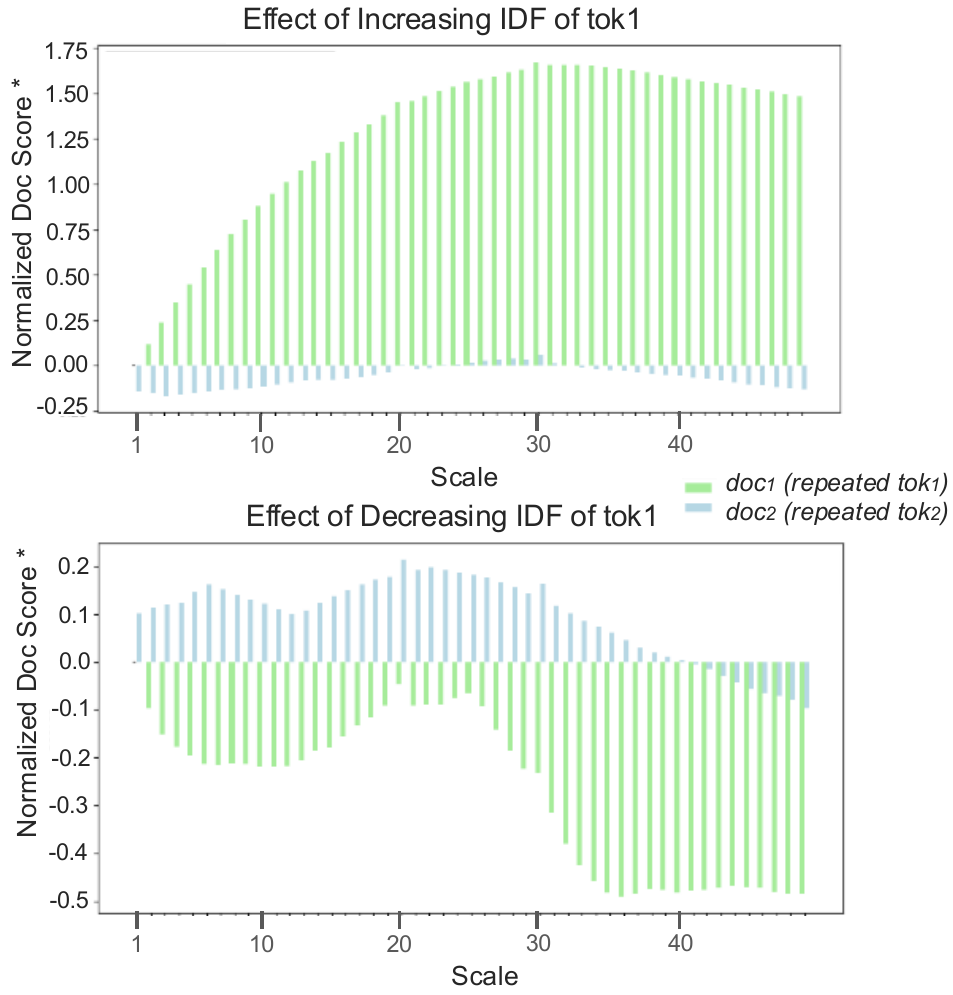}
    \footnotesize{\\ * We normalize the scores so that the average score of the unscaled doc1 is 0.}
    \caption{\textit{Top}: Changes in scores for \texttt{doc1} (repeated \texttt{tok1}) and \texttt{doc2} (repeated \texttt{tok2}) when increasing \texttt{tok1}'s IDF at different scales, averaged across all samples. Increasing \texttt{tok1}’s IDF raises \texttt{doc1}’s score more than \texttt{doc2}’s.  \textit{Bottom}: Decreasing \texttt{tok1}’s IDF produces the inverse effect, with slight non-monotonic deviations suggesting optimal editing windows.}
    \label{fig:model_edit_result}
    \vspace{-1em}
\end{figure}

Figure~\ref{fig:model_edit_result} shows that scaling \texttt{tok1}'s IDF up or down causes a corresponding monotonic increase (or decrease) in \texttt{doc1}'s score, providing causal evidence that the model uses \( U_0 \) to encode IDF and suggests that it sums soft-TF by IDF values as hypothesized in \S4.1 (see additional details about IDF and Relevance Scoring Heads' attention pattern correlation in Appendix~\ref{sec:idf_attention}).

\textbf{Comparison to BM25.}
This completes the BM25 component set: the embeddings encode IDF, and Relevance Scoring Heads modulate the amount of soft-TF extracted by the \texttt{[CLS]} token based on that IDF. This is similar to BM25, where each query term’s TF is weighted by its IDF to prioritize informative terms. In the next section, we formally validate that Relevance Scoring Heads combine soft-TF and IDF in a BM25 manner to compute the final relevance score.

\section{Validation of BM25-like Computation}

\label{sec:circuit-function-approximatio}

To validate whether Relevance Scoring Heads perform a BM25-style computation as hypothesized in \S\ref{sec:relevance_scoring_heads}, we first formalize the hypothesized function of the heads and the information flowing into them as a BM25-style linear function. Next, we evaluate this linear model's ability to reconstruct the cross-encoder scores by examining how well the hypothesized linear model fits the data.

\subsection{Formalizing the Hypothesized Function}
In \S\ref{sec:relevance_scoring_heads}, we hypothesize that Relevance Scoring Heads compute a summation of Matching Scores weighted by IDF. The Matching Score incorporates soft-TF, along with term saturation and document length signals, all of which are components of the BM25 function. If this hypothesis that these components interact in a BM25-like manner holds, then the \textit{Semantic Scoring Circuit} can be expressed as a linear function:

{\small
\begin{equation}
\begin{aligned}
\sum_{i=1}^{N}
&\;\text{linear\_combo}\Bigl(
    -U_0(q_i),\;\text{MS}_{\mathrm{total}}(q_i,d_j),\\
&\quad -\,U_0(q_i)\,\cdot\,\text{MS}_{\mathrm{total}}(q_i,d_j)
\Bigr)
\end{aligned}
\end{equation}
}

where the components of the linear combination are defined as follows:
\begin{enumerate}
    \item \( U_0(q_i) \): The value of \( q_i \) in the \( U_0 \) vector, representing the model's interpretation of \( q_i \)'s IDF.
    \item \( \text{MS}_{\text{total}}(q_i, d_j) \): The total Matching Score, computed as the sum of Matching Scores from individual heads (\( \text{MS}_{H_k} \)) weighted by learned weights \( \alpha_k \). Each Matching Score represents the sum of attention values from \( q_i \) to all document tokens:
    \[
    \text{MS}_{\text{total}}(q_i, d_j) = \sum_{k=1}^{13} \alpha_k \cdot \text{MS}_{H_k}
    \]
    \item \( -U_0(q_i) \cdot \text{MS}_{\text{total}}(q_i, d_j) \): The interaction term, modeling the product of IDF and TF in BM25.
\end{enumerate}

By incorporating both the hypothesized computation function and the earlier components \( U_0 \) and \( MS \), the linear model effectively represents the hypothesized circuit. 

\subsection{\textbf{Assessing the Linear Model Fit}}

We test whether our hypothesis holds by comparing the linear model’s effectiveness against the cross-encoder’s actual relevance scores.

First, we train a linear regression model using our base dataset, limiting queries to five tokens to keep the number of coefficients manageable. For each forward pass, we extract two features for each query token: (1) the \(\text{MS}\) (Matching Score), calculated as the sum of the query token's attention over document tokens from the 13 Matching Heads, and (2) its value along the top singular vector in \(U_0\) of the decomposed embedding matrix. These features form the input \textit{x} to the linear regression model, while the cross-encoder's relevance scores are the target \textit{y}. Finally, we evaluate how well the linear representation of the Semantic Scoring Circuit predicts the cross-encoder's relevance scores with an 80/20 train-test split.

The linear regression model achieves a high Pearson correlation (\textit{r} = 0.8157, \textit{p} < 0.001) with ground-truth relevance scores, showing that it captures the core of the cross-encoder's scoring mechanism in a simplified and interpretable form. This correlation surpasses that of the traditional BM25 scoring function under optimized parameters (\(k=0.5, b=0.9\); corr = 0.4200, \textit{p} < 0.001), which demonstrates that the discovered \( U_0 \) and \( MS \) components effectively capture the signals that the cross-encoder utilizes for ranking. The linear model's strong generalization to unseen datasets and varying query lengths (details in Appendix \ref{sec:linear_generalization}) further confirms that our circuit understanding captures the cross-encoder's core mechanism for relevance computation.

\section{Discussion}

\subsection{A Two-Stage Process for Relevance Computation}
Similar to \citet{tenney2019bert}, who find that BERT rediscovers the classical NLP pipeline, encoding syntax in its lower layers and semantics in its upper, our circuit analysis reveals a corresponding two-stage process. In the early and middle layers, Matching Heads extract lexical signals (e.g., soft-TF matches), reflecting prior work that argues document representations are query-dependent \citep{qiao2019understanding}. In the upper layers, Contextual Query Representation and Relevance Scoring Heads aggregate these signals into BM25-like relevance scores, consistent with \citet{zhan}, who argue that document tokens are largely query-independent. This dual-stage mechanism reconciles these two conflicting viewpoints by showing how query-specific signals migrate from document to query representations across layers and provides a more nuanced understanding of how relevance is computed.

\subsection{Potential Downstream Applications}

In \S\ref{IDF}, we show how model editing can be used to \textit{upscale} or \textit{downscale} term importance by modifying the encoded IDF values in the embeddings, effectively introducing ``tuning dials'' for fine-grained control over model behavior. These insights enable two potential downstream applications.

\subsection{Mitigating Adversarial Attacks}
\label{sec:adv_attack}
\begin{figure}[!ht]
    \centering
    \includegraphics[width=\columnwidth]{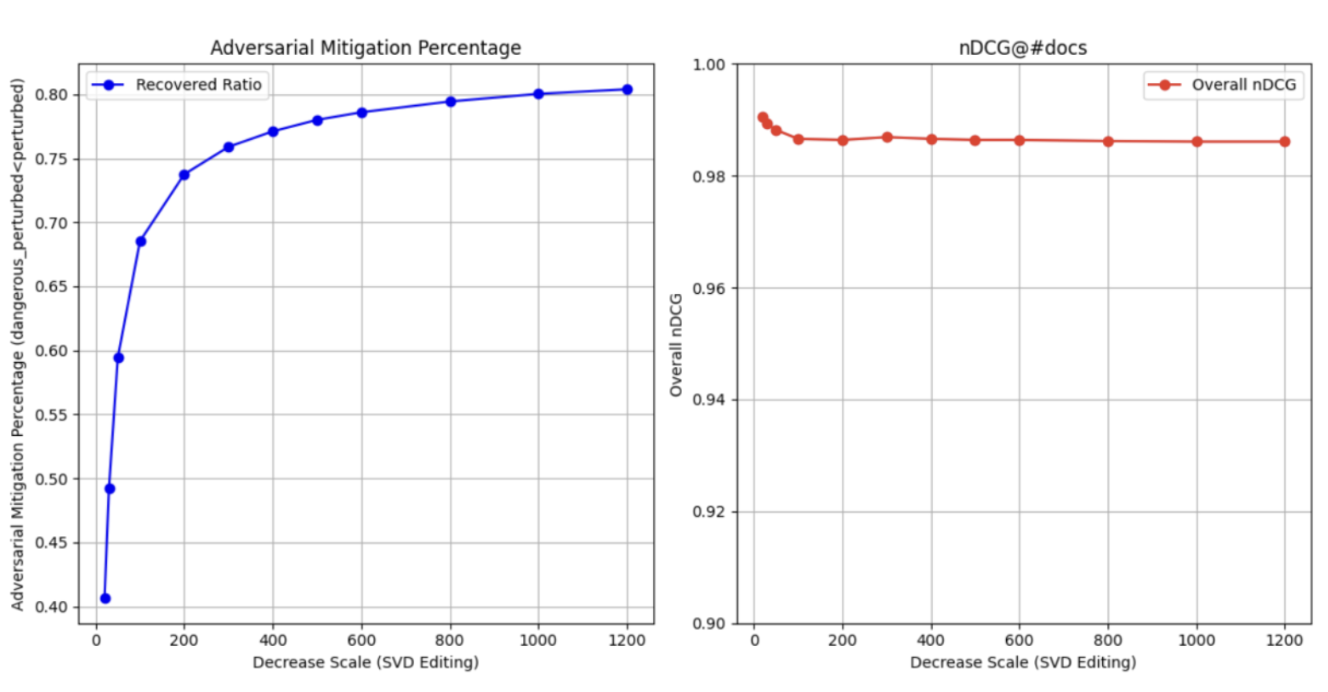}
    \caption{\textit{Left}: Decreasing the IDF of dangerous tokens shows an increasing trend in adversarial mitigation proportion. \textit{Right}: The intervention maintains a high NDCG at all levels, demonstrating that the general ranking capability of the cross-encoder remains unaffected.
}
    \label{fig:ModelEdit_dangerous}
\end{figure}

First, targeted model editing could mitigate unsafe or adversarial content without impairing the model’s ranking effectiveness. 

As an initial test, we construct a dataset using obscene and offensive words~\citep{ldnoobw_badwords}, filtering out multi-word entries and injecting these unsafe tokens into the safe samples of our TFC datasets. Next, we inspect the subgroup of queries and documents where inserting an unsafe token in the document significantly increases the relevance score. Our goal is to ``erase'' the effect of the dangerous token by reducing its importance. 

Among 17,537 adversarial samples, our approach achieves a 80.396\% success rate (i.e., the unsafe document is ranked lower than the safe document). For example, when we downweight the target unsafe term by a large factor (e.g., -1200), the model maintains a high nDCG across all ranks, with a score of 0.9861, which is only a 1.39\% drop in ranking performance. 

These results provide preliminary evidence for the potential of localized editing in the IDF-storing low-rank matrix.

\subsection{Parameter Efficient Fine-Tuning}
\label{sec:peft}

\begin{figure}[!ht]
    \centering
    \includegraphics[width=\columnwidth]{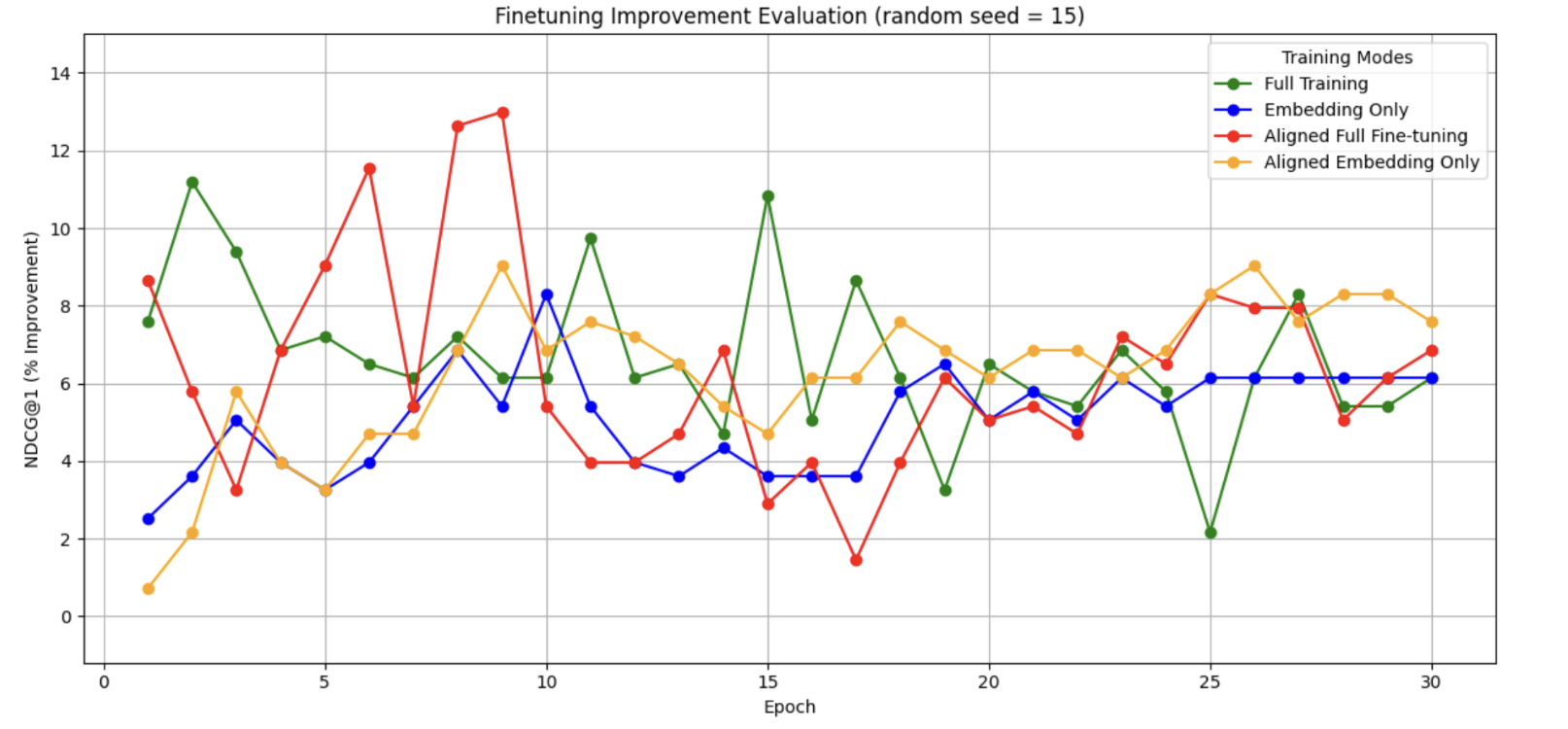}
    \caption{For random seed 15, aligned full fine-tuning achieves the best performance, and aligned embeddings only achieve performance compared to full fine-tuning.}
    \label{fig:ModelEdit_finetune}
\end{figure}

Second, aligning internal representations with ground-truth IDF scores could serve as a powerful initialization strategy for fine-tuning, leading to more efficient retrieval pipelines. In other words, rather than updating the entire parameter space, fine-tuning the embedding matrix to better encode IDF values aligned with the retrieval domain may suffice.

To test this hypothesis with a preliminary experiment, we use the nfcorpus data, a Nutrition Fact retrieval dataset, from BEIR \cite{beir}. We fine-tune on 10 epochs and evaluate performance across four conditions: (A1) full fine-tuning, (A2) IDF-aligned full fine-tuning (aligning token IDFs to the dataset using the model-editing method from \S\ref{IDF}), (B1) embedding-only fine-tuning, and (B2) IDF-aligned embedding-only fine-tuning. 

The results in Figure~\ref{fig:ModelEdit_finetune} across three random seeds show that IDF-aligned full fine-tuning (A2) achieves the highest NDCG@1, while IDF-aligned embedding-only fine-tuning (B2) outperforms unaligned embedding-only fine-tuning (B1). Additionally, full fine-tuning (A1) and embedding-only fine-tuning (B1) show no significant difference, supporting the idea that adaptation primarily refines the embedding matrix rather than the full model weights. 

These findings reveal preliminary indications that fine-tuning just on the embedding matrix could potentially be sufficient to adapt cross-encoders to new domains and reveal IDF-aligned initialization as a possible pathway for efficient adaptation.

% short version of conclusion
\section{Conclusion} 
\label{conclusion}
In this work, we mechanistically uncover the core components of the relevance scoring pathway in a BERT-based IR model, revealing how it leverages contextual representations to implement a semantic variant of BM25. Our fine-grained analysis lays the foundation for applying interventions to build more transparent and controllable models, enabling personalization, bias mitigation, and parameter-efficient adaptation. More broadly, identifying universal components or architecture-specific mechanisms contributes to the interpretability of IR models and informs the design of controllable, efficient ranking systems for real-world deployment.

\section*{Limitations}

In this work, we focus on analyzing the behavior of a single cross-encoder in order to deeply analyze its mechanisms. Future work should investigate the generalizability of this behavior to other neural IR models with the same or different architectural base.

Additionally, while the Semantic Scoring circuit bears a strong resemblance to BM25, there may also be other factors influencing relevance computation that are out of the scope of investigation for this work. In this work, we compare it to BM25 because of its well-known status and the fact that previous research has shown that neural IR encode a BM25-like relevance signal \citep{wang2021bert, rau2022different, yates2021pretrained, macavaney2022abnirml}. However, there are other term-matching models (e.g., TF-IDF, QL, etc.) that also rely on similar retrieval heuristics, and it is possible that the model implements these as well. 

Furthermore, we note that our BM25-like approximation of the Semantic Scoring circuit does not fully approximate the cross-encoder's true relevance scores. There are a couple potential explanations for this: (1) the non-linearity of neural models, which our linear regression model does not capture, (2) potentially unexplored components such as additional attention heads that could be uncovered through a different choice of dataset or multi-layer perception (MLP) layers not covered in our analysis, and (3) the incorporation of signals beyond traditional relevance heuristics, such as real-world knowledge learned during pre-training \cite{macavaney2022abnirml}. We leave the investigation of these avenues for future work.

Finally, although we intend for our analysis to inform future work on efficiency, transparency, and performance improvements, like most research, it is possible that a malicious actor could apply these insights to produce IR models that are robust to adversarial use cases.

\section*{Acknowledgments}

We thank the Health NLP Lab for valuable suggestions and discussions that improved this work.

\bibliography{main}

\clearpage
\appendix

\section{Activation Patching on Attention Heads}

\begin{figure*}
    \centering
    \includegraphics[width=1\linewidth]{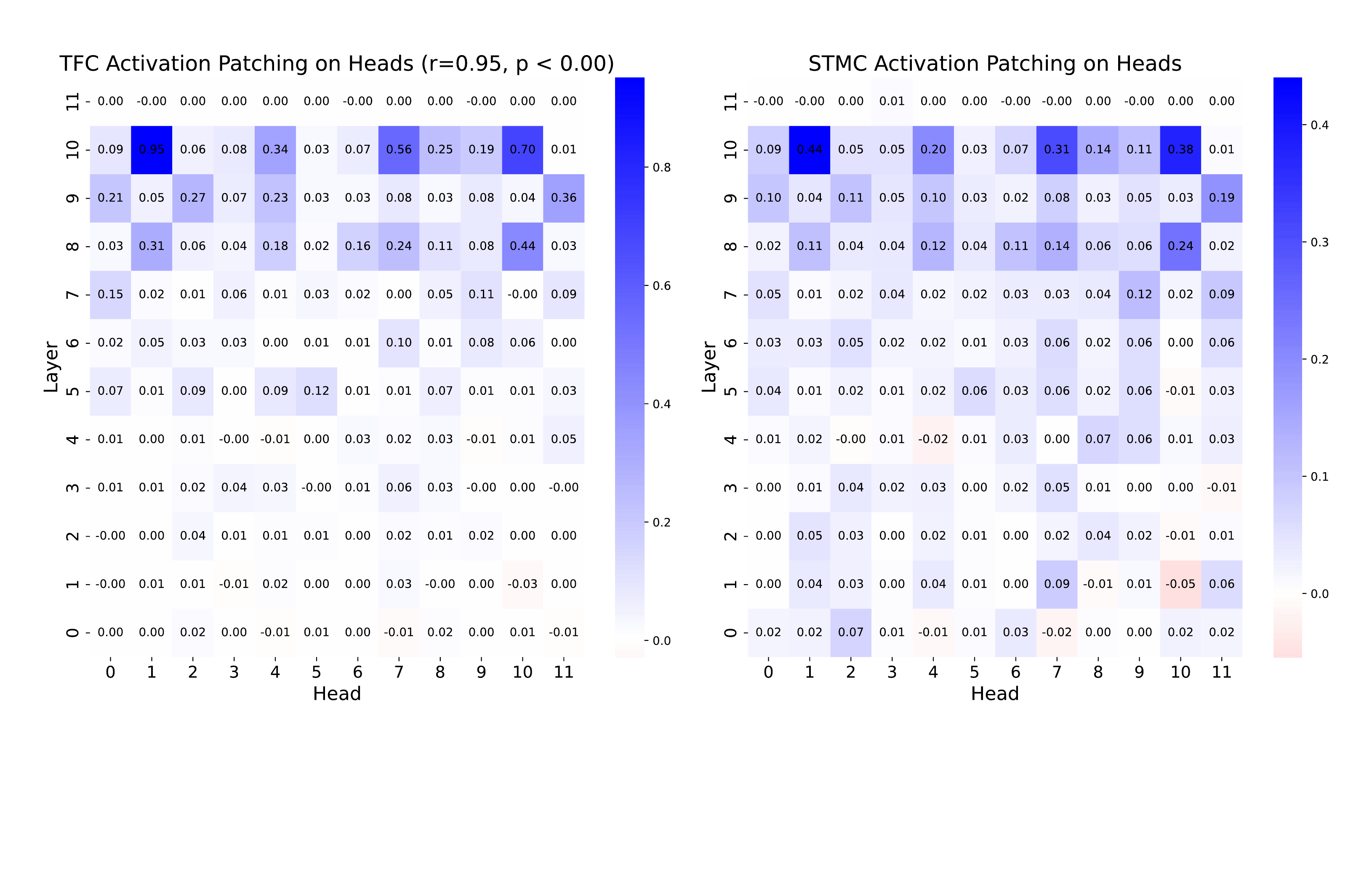}
    \caption{Activation Patching Results for TFC and STMC diagnostic datasets show a correlation of 0.96, which suggests that the model employs highly similar components for exact and soft matches.}
    \label{fig:appendix_activation_patching}
\end{figure*}

Prior to path patching, we conduct an exploratory analysis to compare the similarity between exact and semantic matches. Figure~\ref{fig:appendix_activation_patching} shows that the patching effect has a high correlation of 0.96, suggesting that the model employs highly similar components for both exact and soft matches. These initial results on the indirect effect of these components provide a foundation to investigate their direct effects further with path patching.

\section{Example of Semantic Scoring Circuit}
\label{sec:example_of_semantic_scoring_circuit}
\begin{figure*}
    \centering
    \includegraphics[width=0.78\linewidth]{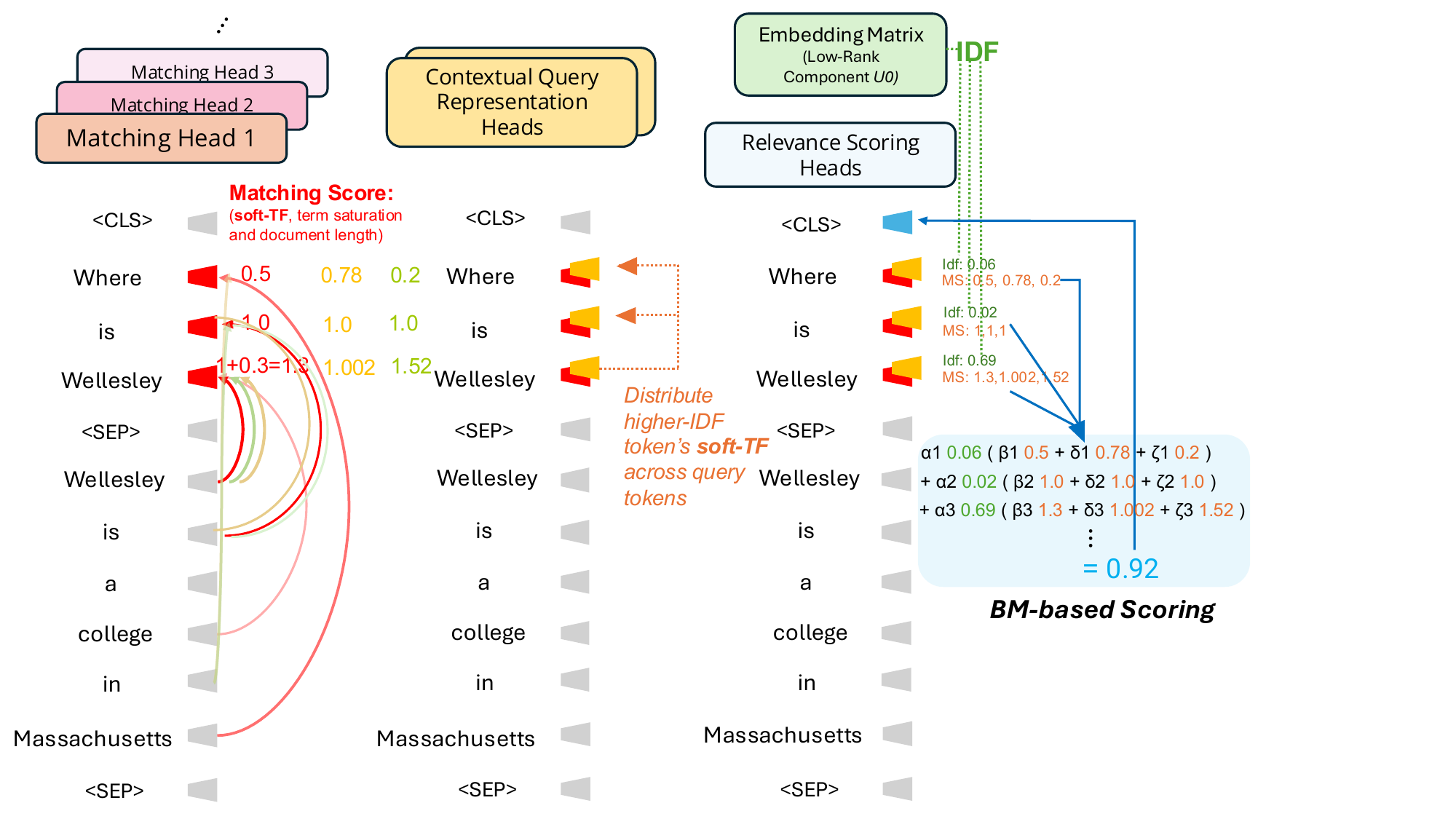}
    \caption{Example walkthrough of the hypothesized circuit. We find that the cross-encoder rediscovers a semantic variant of BM25. Specifically, Matching Heads, Contextual Query Representation Heads, and the embedding matrix compute, process and send BM25-like components’ information to the Relevance Scoring Heads which finally compute the relevance score in a BM25-like manner. The trapezoid represents the residual stream representation for each token position.}
    \label{fig:pipeline-example-overview}
\end{figure*}

Figure~\ref{fig:pipeline-example-overview} shows an example of the hypothesized semantic scoring circuit.

In the figure, the model receives an example input:
``\texttt{[CLS] Where is Wellesley [SEP] Wellesley is a college in Massachusetts. [SEP]}``.

(1) Matching Heads: These heads generate $MS(\texttt{where})$, $MS(\texttt{is})$, and $MS(\texttt{wellesley})$ that mainly capture query term soft-TF values. For example, the query token \texttt{wellesley} strongly attends to \texttt{wellesley} in the document, and more weakly to related terms like \texttt{wellesley} and \texttt{college}. This attention pattern forms a Matching Score $MS(\texttt{wellesley})$ that also reflects document length normalization (the longer the document, the smaller the soft $MS(\texttt{wellesley})$) and term saturation effects (the more \texttt{wellesley} exists in the document, the less additional gain in $MS$).

(2) Query Contextualization Heads: These heads redistribute the soft-TF signal from key content words to surrounding query tokens. For instance, \texttt{where} and \texttt{is} may receive part of \texttt{wellesley}'s soft-TF signal, allowing the model to amplify high-IDF query terms’ soft-TF signals.

(3) IDF storage: The IDF of all query terms is stored in a dominant low-rank vector of the model’s embedding matrix.

(4) Relevance Scoring Heads: Finally, the model aggregates the reweighted soft-TF signals across query tokens using a mechanism similar to BM25.

%Here, $Z$ accounts for document length and saturation effects.

\section{Detailed Path Patching Methodology}
\label{sec:path-patching-method}
\begin{figure*}
    \centering
    \includegraphics[width=0.78\linewidth]{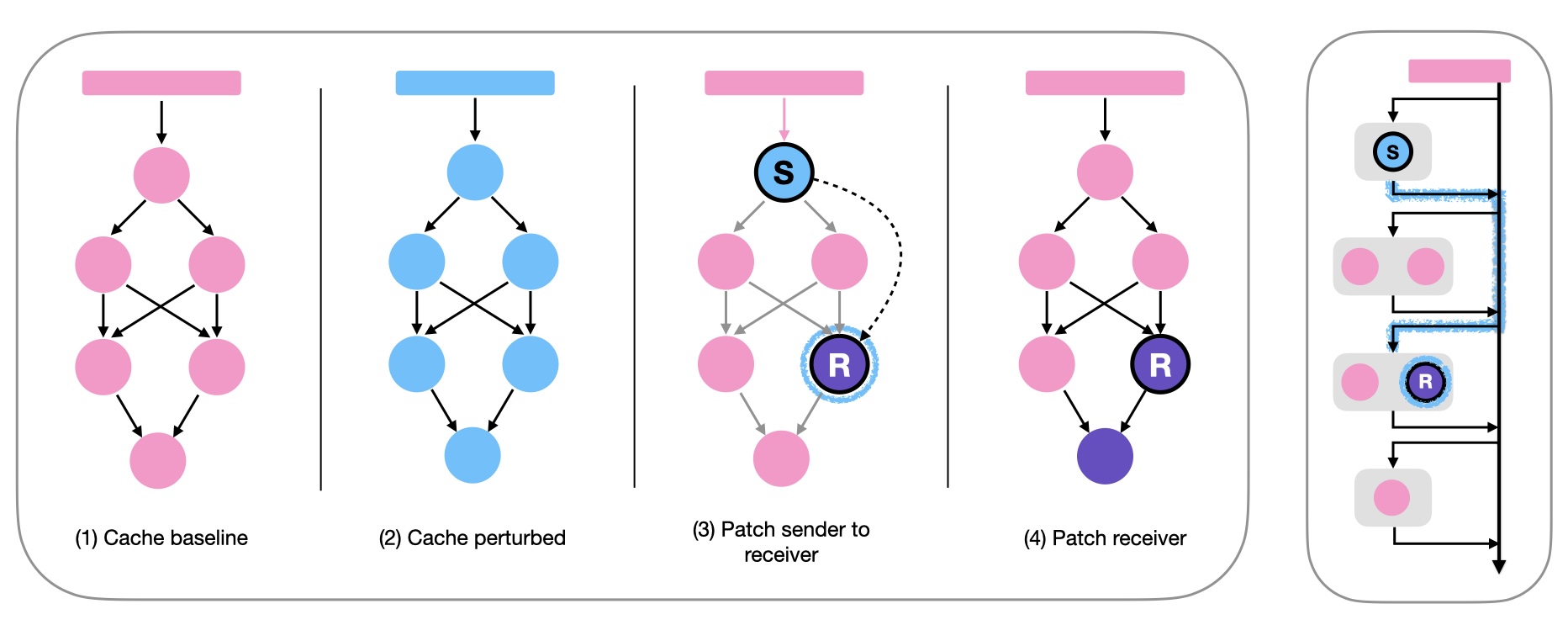}
    \caption{Path patching methodology. \textit{Left:} There are four forward passes: (1,2) Run model on baseline and perturbed inputs and cache activations. (3) To measure the effect of an upstream sender component (\textit{S}) on a downstream receiver component (\textit{R}), run the model on the baseline input, patch in \textit{S}, freeze all other components, and cache the activation of \textit{R}. (4) Run the model on the baseline input and patch in \textit{R}. \textit{Right:} Alternative visualization of step (3) using the residual stream. By allowing only the downstream receiver \textit{R} to be recomputed when the sender \textit{S} is patched, we effectively isolate the direct path from \textit{S} to \textit{R}, while preserving all other paths to \textit{R} as they were in the baseline run.}
    \label{fig:path-patching-overview}
\end{figure*}

The path patching algorithm is an iterative backward process that starts with identifying which upstream components send important information to the logits (Figure~\ref{fig:path-patching-overview}). Specifically, it involves four forward passes through the model to identify which upstream (sender) components send information to the target (receiver) component (i.e., logits): (1) Run on the baseline input \( x_b \) and cache activations. (2) Run on the perturbed input \( x_p \) and cache activations. (3) Select the sender set \( s \), the components whose activations are patched in, and the receiver set \( r \), the components where the effect of patching \( s \) is analyzed. Run the forward pass on \( x_b \), patch in s, freeze all other activations, and recompute \( r' \), which is the same as \( r\) from the \( x_b \) run except for the direct influence from s to r. Cache \( r' \). (4) Run the model on \( x_b \), and patch in \( r' \) values. Measure the difference in logits between \( x_b \) and \( x_p \) to quantify the effect of \( s \) on  \( r\) in terms of passing the additional signal.

The effect of a patch is measured by the difference in logits (which in the case of cross-encoders, is equivalent to the difference in relevance scores). This algorithm is then iteratively repeated for each important upstream component. For more information, we refer the reader to~\citet{wang_ioi} and~\citet{goldowskydill_ppatching}.

We begin by applying path patching to track the path of term-matching, the most fundamental component of BM25. To carry this out, we use the TFC1 and STMC1 diagnostic datasets to identify the components responsible for encoding the relevance signal of an additional query term in a document, providing the foundation for our analysis of model behavior.

\section{Detailed Path Patching Results}
\label{sec:appendix_detailed_ppatch}

\subsection{Path Patching to Final Logits and Intermediate Heads}
\label{sec:ppatch_logits_intermediate}

We perform path patching on 2000 random query and document pairs from the STMC1 and TFC1 diagnostic datasets and collect the results.

\begin{figure*}
    \centering
    \includegraphics[width=1\linewidth]{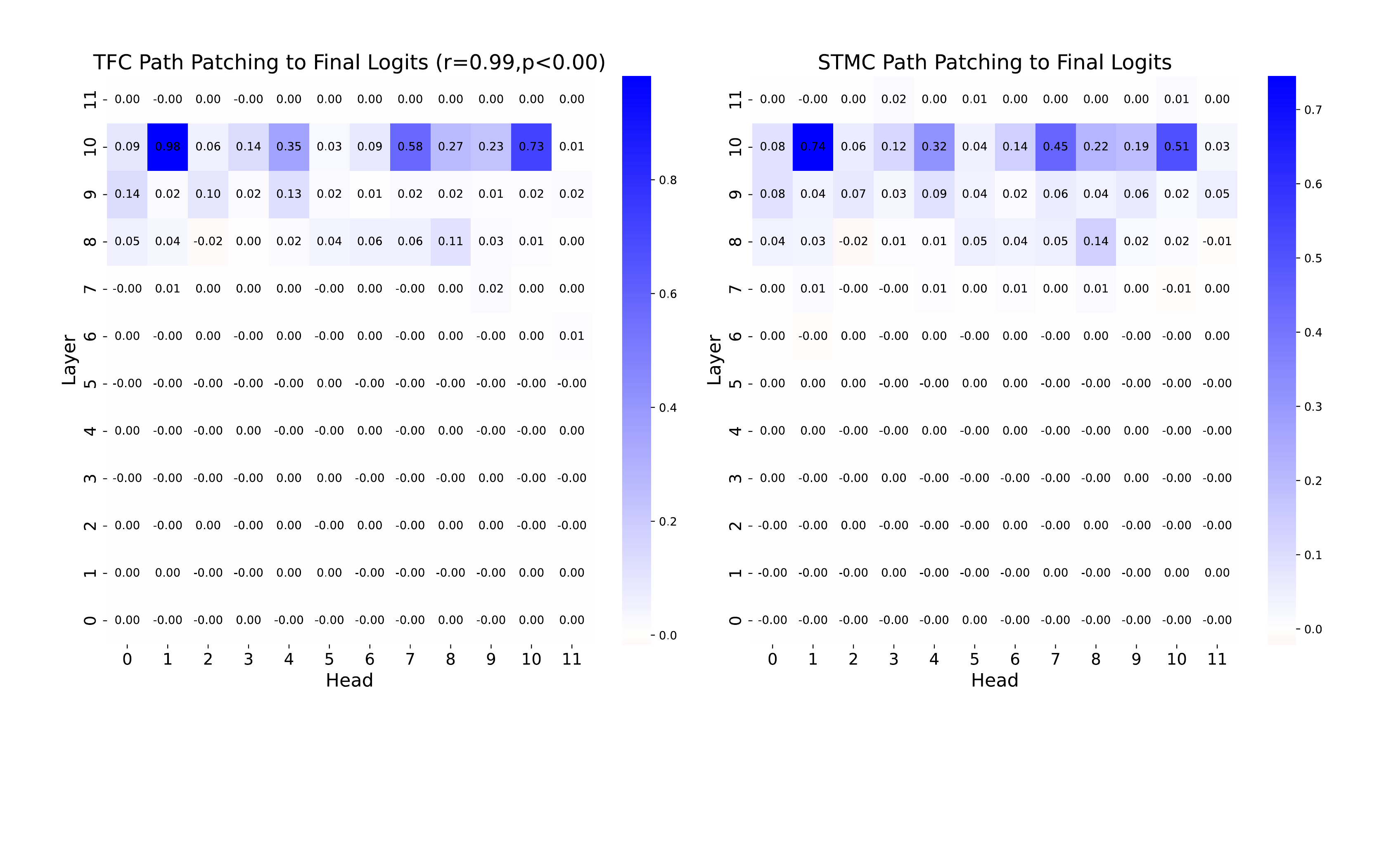}
    \caption{Path Patching results from all attention heads to the final logits. From this, we identify that 10.1, 10.4, 10.7, 10.19, those causing > 30 \%  increase in ranking score, as most significant heads and focus our analysis on these four heads. }
    \label{fig:appendix_path_patching_to_final_logits}
\end{figure*}

\begin{figure*}
    \centering
    \includegraphics[width=1\linewidth]{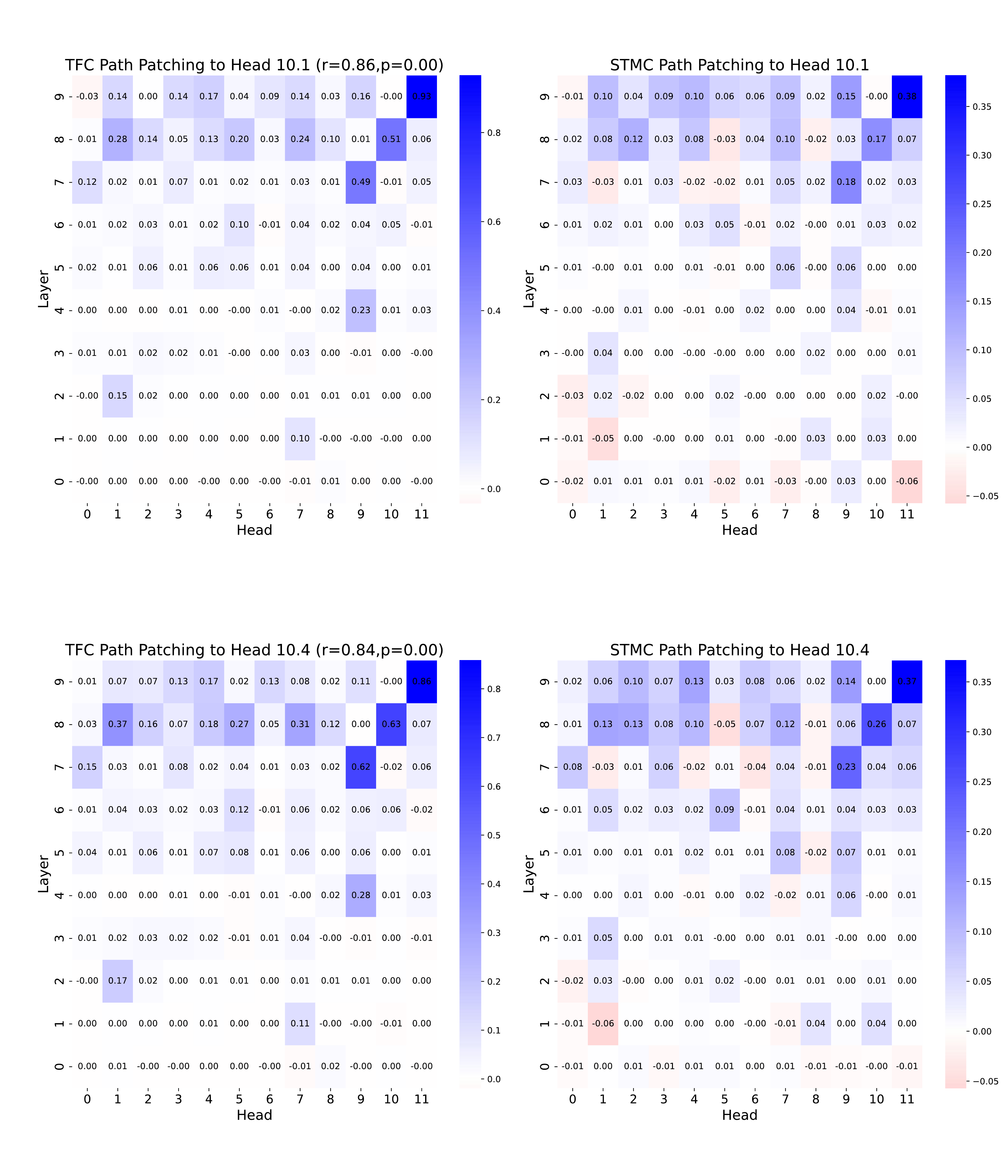}
    \caption{Path Patching results from upstream attention heads to Relevance Scoring Heads. }
    \label{fig:appendix_path_patching_to_rsh1}
\end{figure*}

\begin{figure*}
    \centering
    \includegraphics[width=1\linewidth]{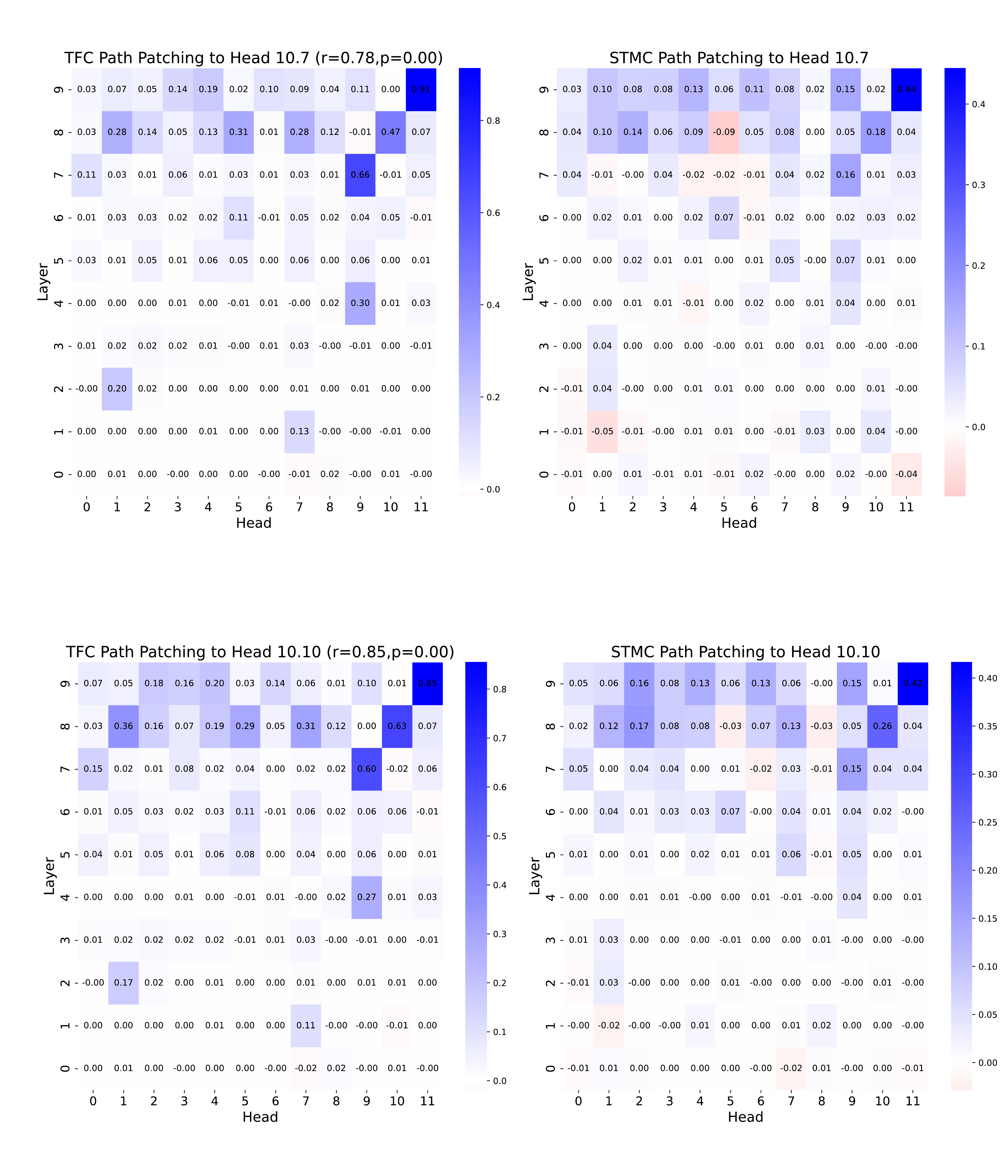}
    \caption{Path Patching results from upstream attention heads to Relevance Scoring Heads. }
    \label{fig:appendix_path_patching_to_rsh2}
\end{figure*}

In this section, we present the results of path patching on TFC1 and STMC1. First, we patch to the final logits and identify the shared Relevance Scoring Heads (Figure~\ref{fig:appendix_path_patching_to_final_logits}). Next, we patch to these heads and discover the shared Query Contextualization Heads (Figures~\ref{fig:appendix_path_patching_to_rsh1},~\ref{fig:appendix_path_patching_to_rsh2}). Finally, by patching to the Query Contextualization Heads, we identify the shared Matching Heads (Figure~\ref{fig:appendix_path_patching_to_qch}).

All resulting heatmaps are located at the end of the appendix for clearer flow for the other appendix sections.

\subsection{Path Patching to Matching Heads}
\label{sec:ppatch_matching_heads}
\begin{figure*}
    \centering
    \includegraphics[width=0.78\linewidth]{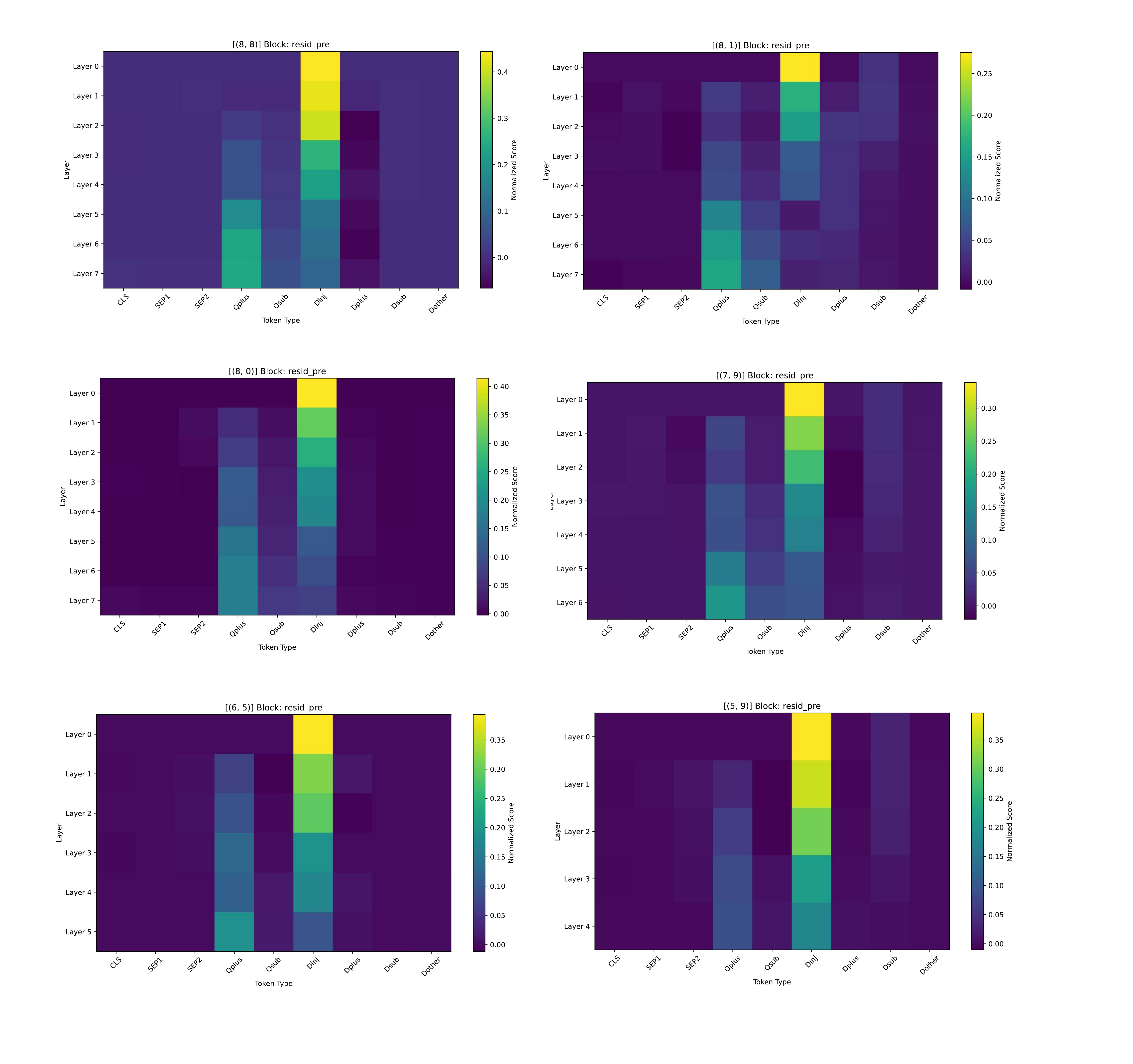}
    \caption{TFC1 path patching into Matching Heads (layers 8.8, 8.1, 8.0, 7.9, 6.5, 5.9). The residual stream values are patched, with lighter regions indicating stronger path-patching contributions (i.e., where the TFC1 signal flows), and darker regions indicating weaker contributions.}
    \label{fig:ppatch_MS_tfc1}
\end{figure*}

\begin{figure*}
    \centering
    \includegraphics[width=0.78\linewidth]{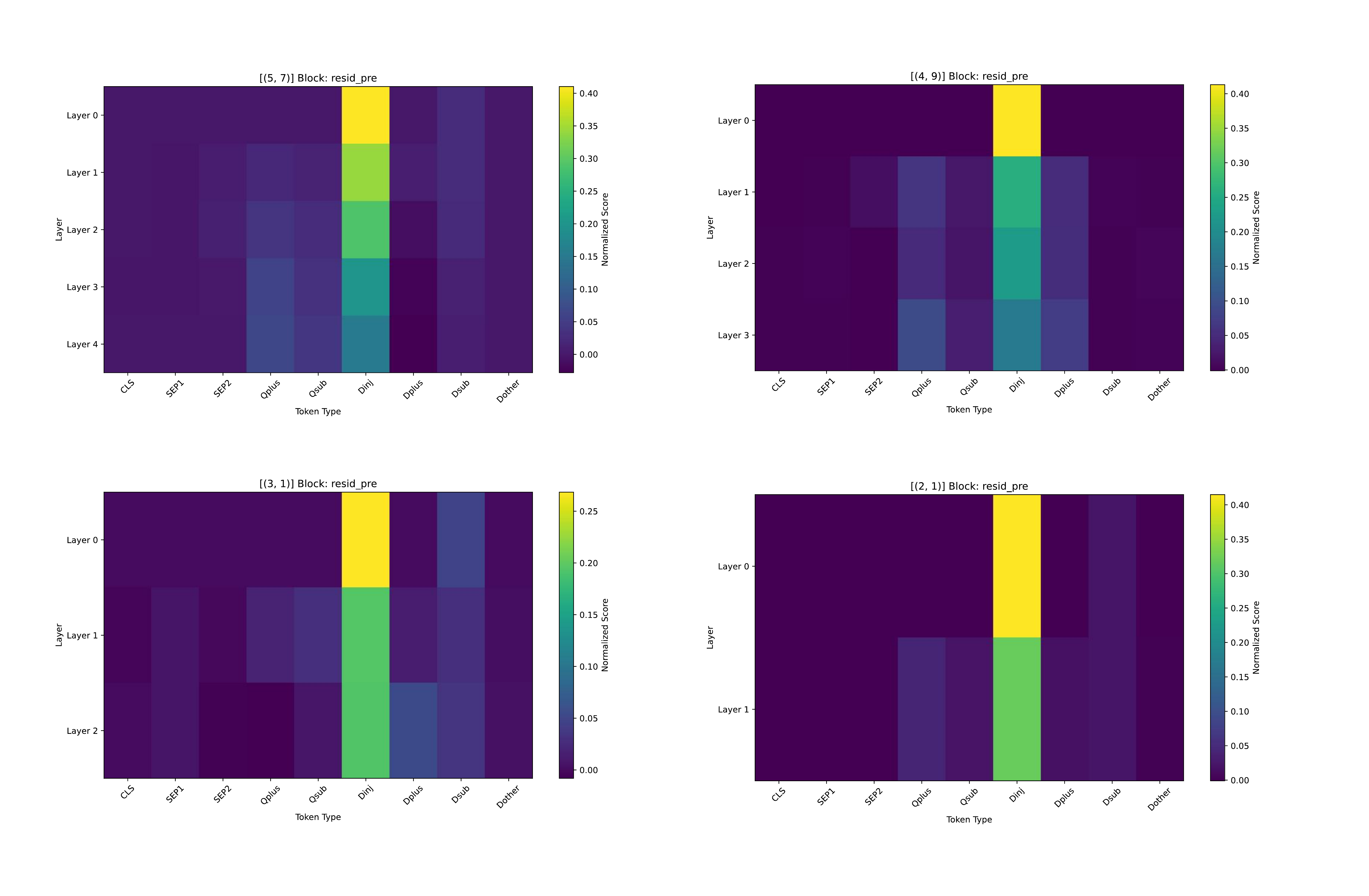}
    \caption{TFC1 Path Patching to Matching Heads (5.7, 4.9, 3.1, 2.1)}
    \label{fig:ppatch_MS_tfc2}
\end{figure*}

\begin{figure*}
    \centering
    \includegraphics[width=0.78\linewidth]{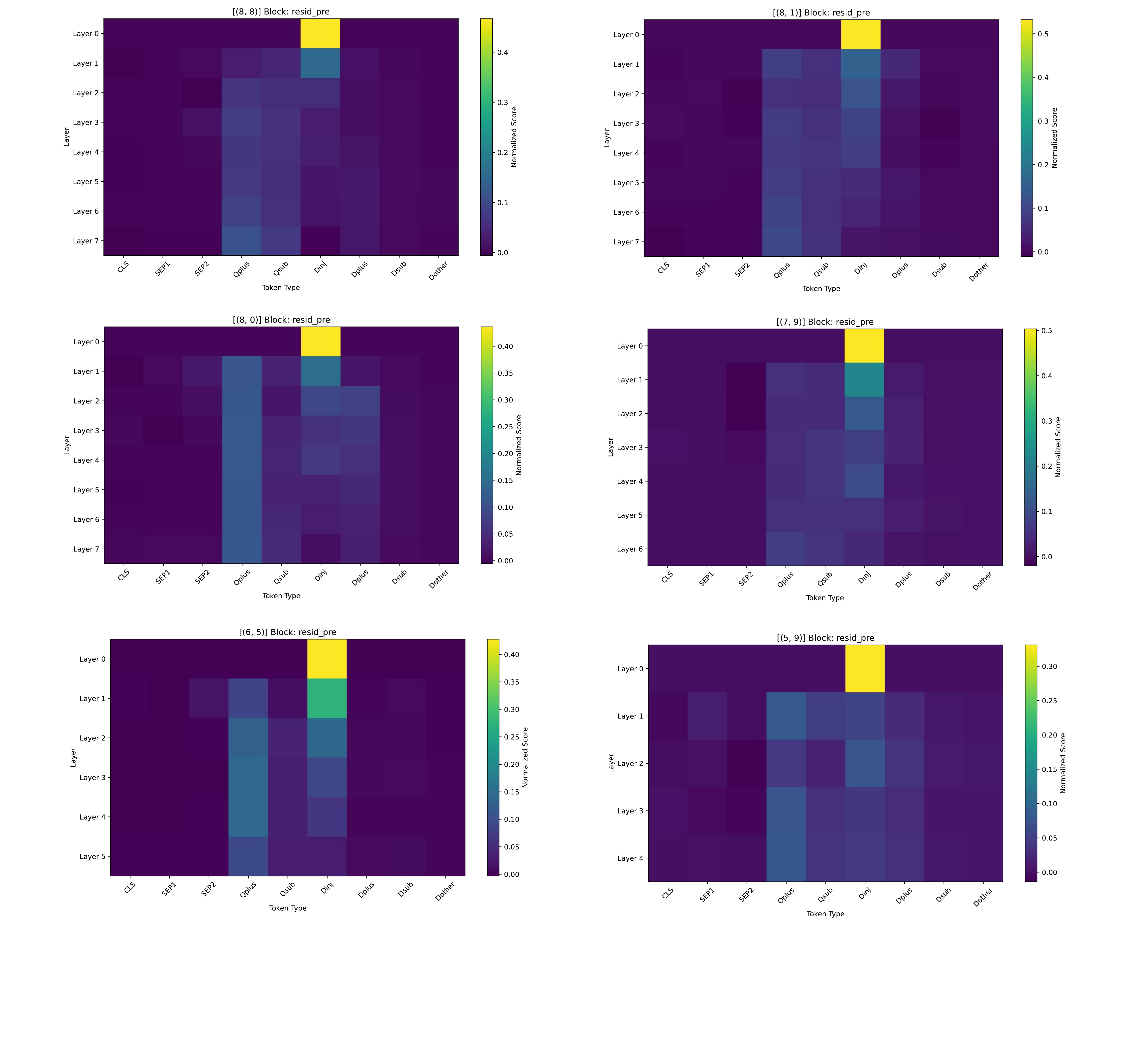}
    \caption{STMC1 Path Patching to Matching Heads (8.8, 8.1, 8.0, 7.9, 6.5, 5.9,)}
    \label{fig:ppatch_MS_stmc1}
\end{figure*}

\begin{figure*}
    \centering
    \includegraphics[width=0.78\linewidth]{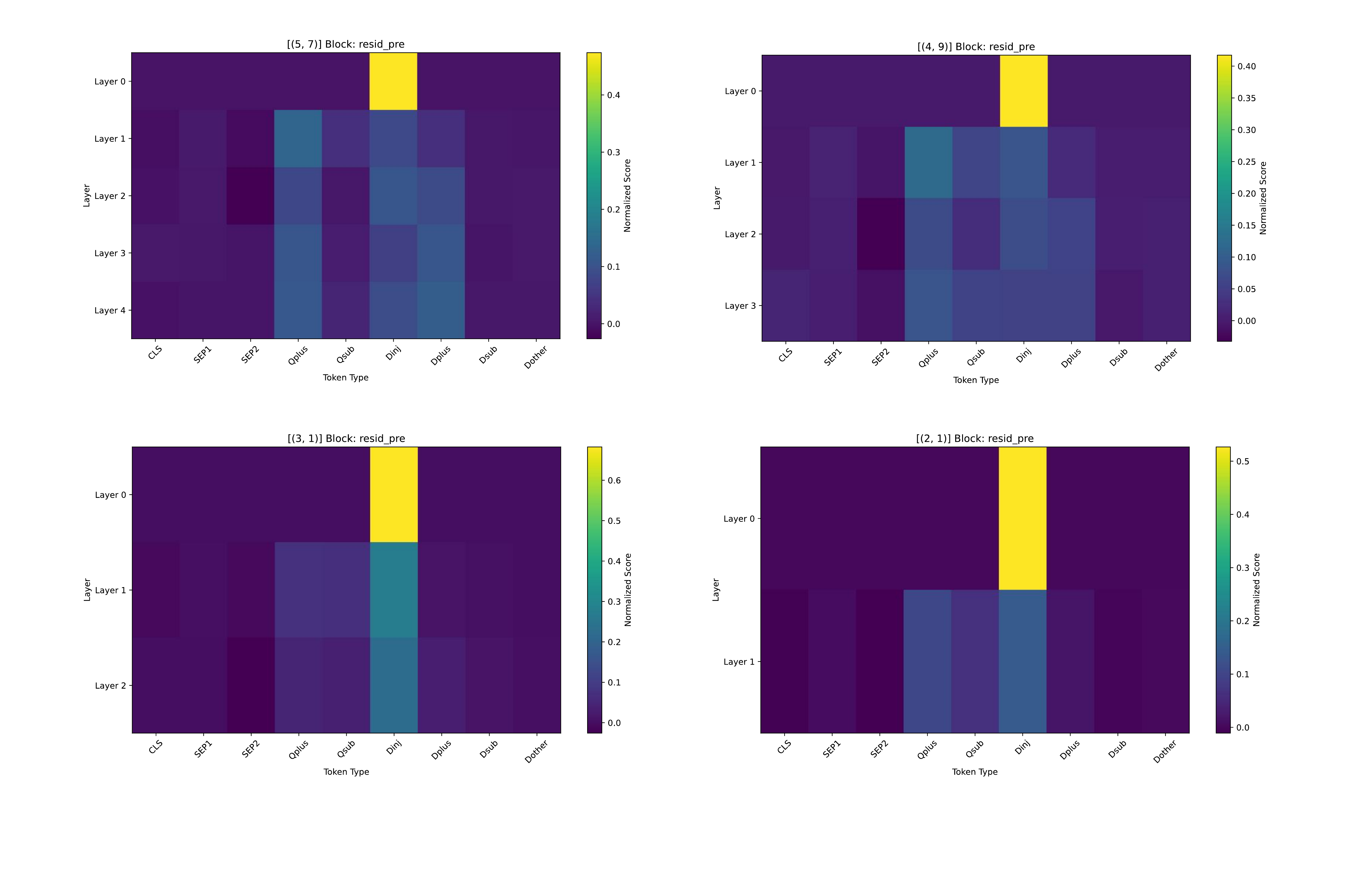}
    \caption{STMC1 Path Patching to Matching Heads (5.7, 4.9, 3.1, 2.1)}
    \label{fig:ppatch_MS_stmc2}
\end{figure*}

We path patch the residual stream (resid\_pre) on each of the Matching Heads using 100 random pairs from the TFC1 and STMC1 datasets (Figures \ref{fig:ppatch_MS_tfc1}, \ref{fig:ppatch_MS_tfc2}, \ref{fig:ppatch_MS_stmc1}, \ref{fig:ppatch_MS_stmc2}). Path patching to the Matching Heads, aimed at tracking the flow of soft-TF information, reveals that at the beginning of the residual stream at layer 0, which corresponds to the embeddings, has the most direct influence on the Matching Heads. In contrast, there is minimal impact from both the attention and MLP layers. This supports the conclusion that the Matching Heads are the primary ``generator'' of soft-TF, attending to soft matches based on semantic similarity from the embeddings. Thus, we conclude our tracking of soft-TF information, with the Matching Heads identified as the most upstream heads.

\begin{figure}[!ht]
    \centering
    \includegraphics[width=0.45\textwidth]{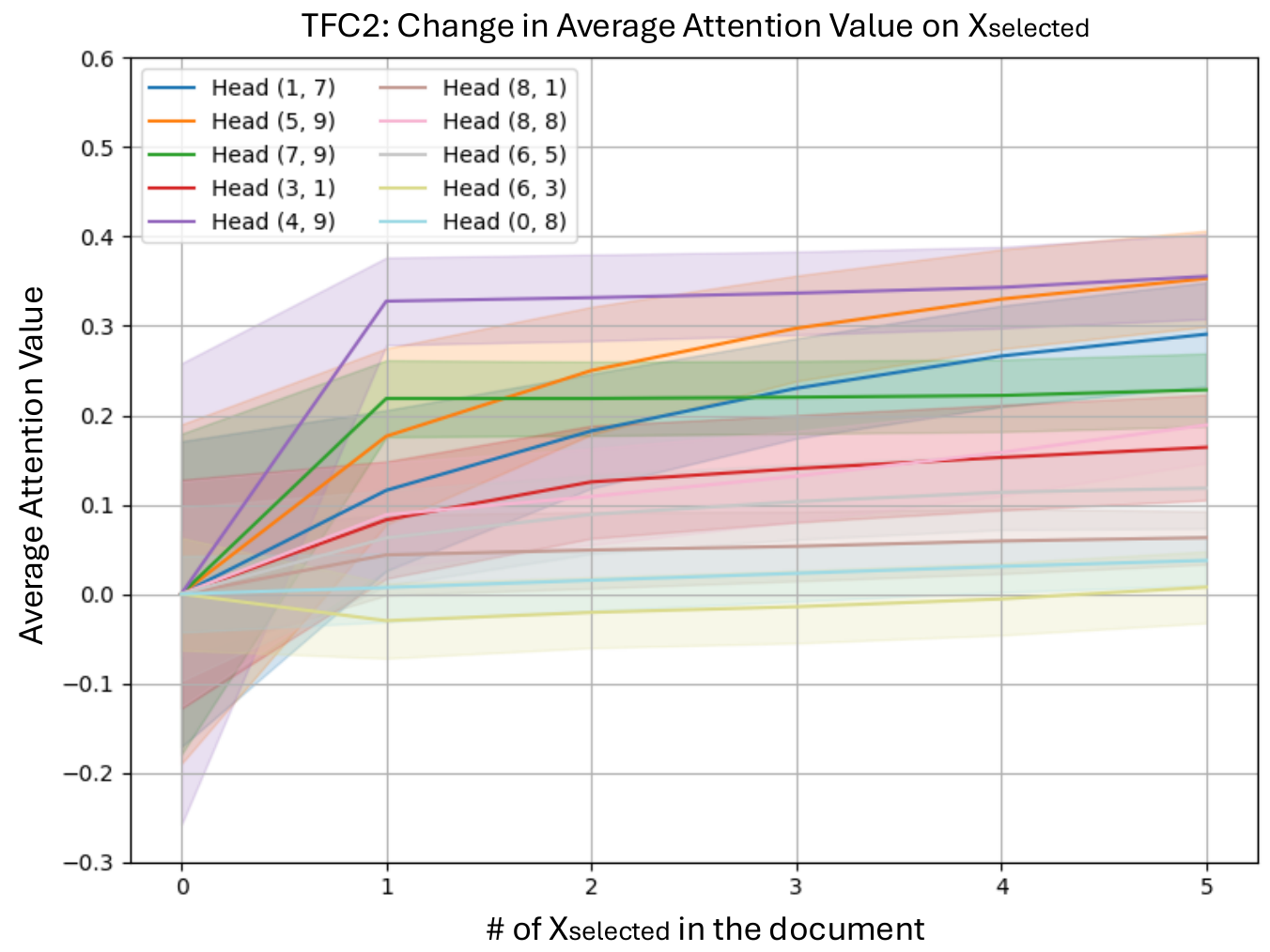}
    \caption{For the majority of Matching Heads, after the initial occurrence of \( X_\text{selected} \), which causes a sharp increase in attention, subsequent occurrences result in only minimal incremental increases. This aligns with TFC2, which states that additional term occurrences yield smaller improvements.}

    \label{fig:MH_other_signals_tfcs}
\end{figure}

\begin{figure}[!ht]
    \centering
    \includegraphics[width=0.45\textwidth]{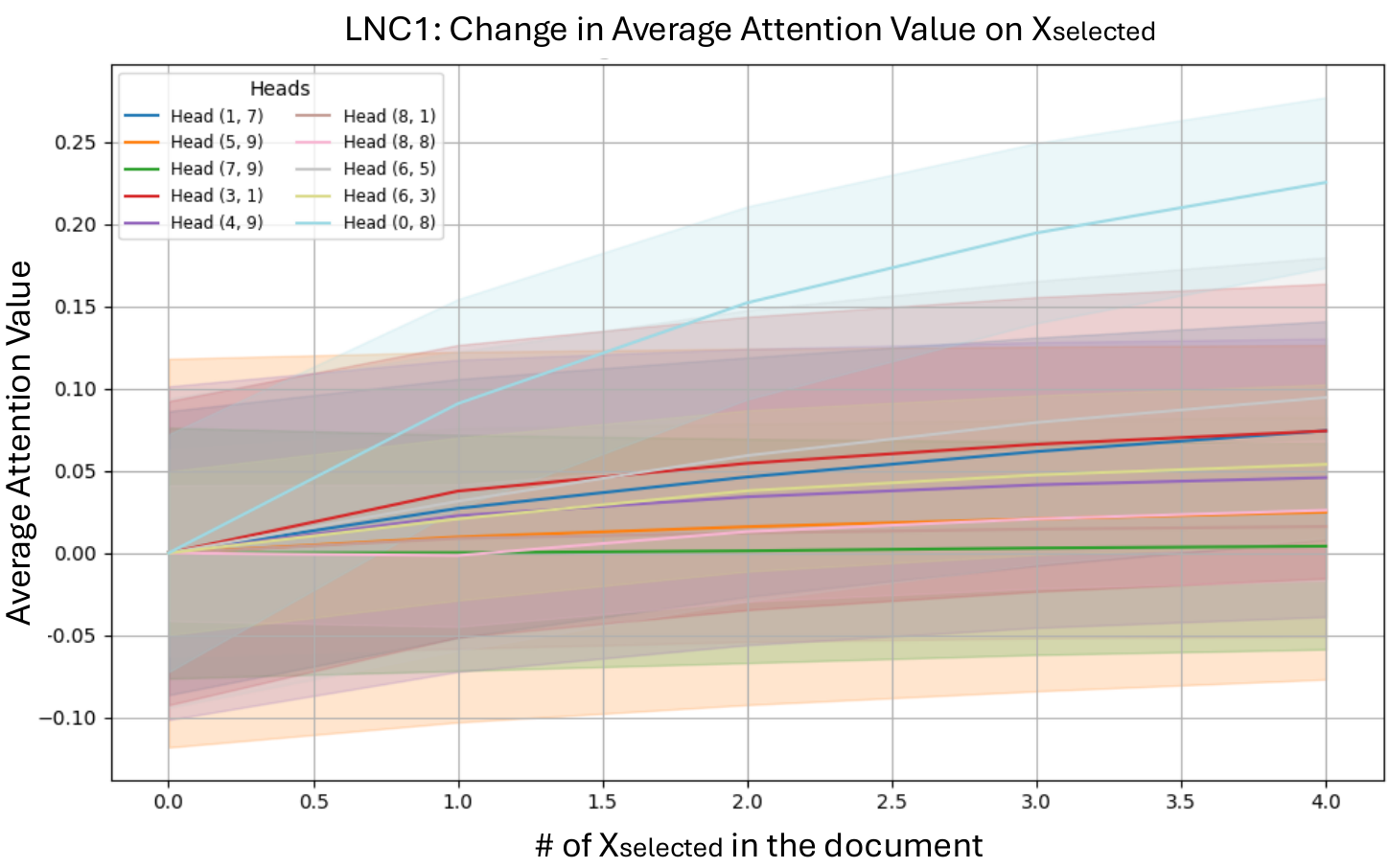}
    % \caption{This figure shows attention value trend for the selected terms, but the attention change on unselected terms exhibits a very similar trend: longer documents generally increase the average attention values assigned to all document tokens regardless of whether they match a query token. }
    \caption{Attention trends for selected terms are shown, with unselected terms following a similar pattern: longer documents generally increase the average attention assigned to all document tokens, regardless of whether they match a query token.}
    \label{fig:Discussion_doclen_attn}
    \vspace{-1em}
\end{figure}

\section{Additional Details on Matching Heads and Additional Signals}
\label{sec:matching_heads_additional_signals}

In this section, we provide additional details on the analysis presented in \S\ref{matching_heads}.

\textbf{Matching Score Contains Term Saturation and Document Length Signals.}
\label{sec:term_saturation_doc_length}
The average soft-TF correlation score for Matching Heads (0.500) indicates that their attention values from query tokens to document tokens capture not only semantic similarity but also additional signals. To examine whether these attention values capture term saturation or document length effects, we analyze their behavior under using two diagnostic datasets: (1) TFC2 (increasing occurrences of a query term), and (2) LNC1 dataset (increasing number of irrelevant sentences to simulate the isolated effect of longer document length). 

\paragraph{Saturation Test (TFC2).} We define \(x_{\text{selected}}\) as the selected query term, which is the duplicate term in TFC2 samples shown in Table \ref{tab:axioms_table}. Similarly, \(x_{\text{others}}\) refers to document tokens that are not duplicates of any query token. For each case, we calculate the normalized sum of attention values from all query tokens to \(x_{\text{selected}}\) and \(x_{\text{others}}\), representing the ``total semantic match'' of the matched token or the average across unmatched tokens, respectively. 

We hypothesize that if the Matching Score contains the term saturation signal, then given more duplicate query terms (TFC2), the summed attention for that term will rise sharply at first and then plateau. As the occurrences of \(x_{\text{selected}}\) increase, its average attention value grows (Figure~\ref{fig:MH_other_signals_tfcs}). On the other hand, the attention for \(x_{\text{others}}\) remains relatively constant, which aligns with our soft-TF understanding: only document tokens that are semantically similar to a query token get nonzero attention values, proportional to the extent and frequency of semantic similarity. 

\begin{table*}[t]
  \centering
  \begin{tabular}{lcc}
    \toprule
    \textbf{Model Pair} & \textbf{Median} & \textbf{Mean $\pm$ SD} \\
    \midrule
    \multicolumn{3}{l}{\textbf{Pearson Correlation}} \\
    Cross-encoder \& SemanticBM & \textbf{0.8401} & 0.8301 $\pm$ 0.0314 \\
    Cross-encoder \& Random Features & 0.0005 & 0.0003 $\pm$ 0.0147 \\
    Cross-encoder \& BM25 & 0.4570 & 0.4609 $\pm$ 0.0366 \\
    \midrule
    \multicolumn{3}{l}{\textbf{Spearman Rank Correlation}} \\
    Cross-encoder \& SemanticBM & \textbf{0.7619} & 0.6629 $\pm$ 0.3137 \\
    Cross-encoder \& Random Features & 0.0000 & 0.0011 $\pm$ 0.3844 \\
    Cross-encoder \& BM25 & 0.4643 & 0.3912 $\pm$ 0.3992 \\
    \midrule
    \multicolumn{3}{l}{\textbf{NDCG@10}} \\
    Cross-encoder & 0.5000 & 0.5097 $\pm$ 0.4110 \\
    SemanticBM & \textbf{0.4307} & 0.4511 $\pm$ 0.3769 \\
    BM25 & 0.5000 & 0.5269 $\pm$ 0.4196 \\
    Random Features & 0.3562 & 0.3498 $\pm$ 0.2986 \\
    \bottomrule
  \end{tabular}
    \caption{Comparison of Pearson, Spearman Rank, and NDCG@10 for SemanticBM and Baselines.}
  \label{tab:table_validation}
\end{table*}

\paragraph{Length Test (LNC1).} We hypothesize that if the Matching Score contains the term saturation signal, given irrelevant sentences appended (LNC1), the overall attention to the query’s matches will fall. As expected, injecting more irrelevant sentences leads to an increase in the attention value of all query tokens, regardless of whether they match specific query terms (Figure \ref{fig:Discussion_doclen_attn}). Notably, this \textit{increase} in attention occurs even as the overall relevance score \textit{decreases} (as expected from LNC1), suggesting that attention values encode mixed and composite signals of soft-TF, term saturation, and document length, which the model has learned to disentangle to produce appropriate relevance score changes. The varying degrees of influence on the heads' attention values shown in Figure~\ref{fig:MH_other_signals_tfcs} and~\ref{fig:Discussion_doclen_attn} suggest that some Matching Heads play a more significant role in regulating these effects, and they may collectively approximate the ideal term effects. Thus, we define this complex attention value as the \textbf{\textit{Matching Score}} to reflect that it encapsulates soft-TF, term saturation, and document length signals, three important signals of BM25.

\textbf{Further Patching Experiment to Confirm Matching Heads Are Not Just Writing a Binary Signal of Existence.}
The signals written by most Matching Heads increase as the number of duplicate tokens rises, rather than remaining constant, suggesting that these signals are discrete rather than binary. To demonstrate this, we design an activation patching experiment (\S\ref{matching_heads}) that patches only the Matching Heads using the TFC2 Diagnostic Dataset. Figure~\ref{fig:MH_other_signals_tfcs} shows that for most Matching Heads, increasing the number of duplicate tokens results in a monotonically increasing pattern in their output signals. This behavior further supports our conclusion that Matching Heads compute and encode soft-TF signals rather than merely relaying binary information. 

\section{Additional Details on BM25-like Computation}

In this section, we provide additional details on the Relevance Scoring Heads and the Semantic Scoring Hypothesis in \S\ref{sec:relevance_scoring_heads} and \S\ref{sec:circuit-function-approximatio}.
\subsection{Correlation Between IDF Values and Attention Distribution}
\label{sec:idf_attention}
\begin{figure}[!ht]
    \centering
    \includegraphics[width=0.35\textwidth]{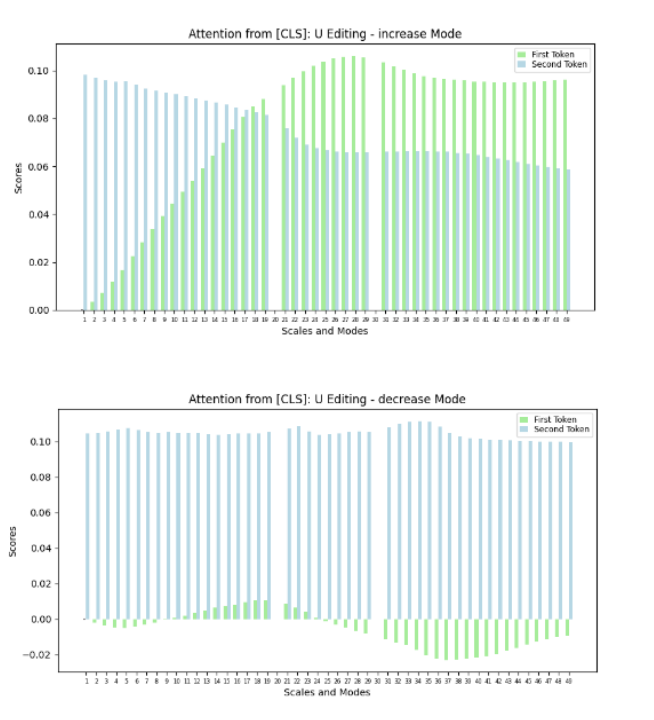}
    
    \caption{\textit{Top}: Changes in attention at 10.1 from \texttt{[CLS]} to \texttt{tok1} when increasing \texttt{tok1}'s IDF at different scales, averaged across all samples. Increasing \texttt{tok1}’s IDF raises attention from \texttt{[CLS]} to \texttt{tok1}, thereby increasing the final score. \textit{Bottom}: Decreasing \texttt{tok1}’s IDF produces the inverse effect, with slight nonmonotonic deviations suggesting optimal editing windows.}
    \label{fig:model_edit_result_attention}
    \vspace{-1em}
\end{figure}

In Section~\ref{sec:relevance_scoring_heads}, we hypothesize that Relevance Scoring Heads combine IDF and soft-TF information to compute final relevance scores in a BM25-like manner. Specifically:  
(A) increasing or decreasing a token’s IDF leads to  
(B) the Relevance Scoring Head 10.1\footnote{We focus on the attention pattern of 10.1 because it is most positively correlated with IDF (\S\ref{sec:relevance_scoring_heads}) and has highest path-patching value.} allocating more or less attention from \texttt{[CLS]} to the token, which  
(C) increases or decreases to the weighting of the token’s soft-TF and thereby the final relevance score.  
\paragraph{A $\rightarrow$ B.} Using the same IDF-scaling setup as in \S\ref{IDF}, we measure the average attention weight from Head~10.1’s \texttt{[CLS]} token to the manipulated token. The resulting curve closely mirrors Figure~\ref{fig:model_edit_result}: tokens with higher IDF receive proportionally more attention, while those with lower IDF receive less. This directly demonstrates the A$\rightarrow$B link. 

\paragraph{A $\rightarrow$ B $\rightarrow$ C.} In Section~\ref{IDF}, we modulate IDF by editing the first SVD component of the embedding matrix. Figure~\ref{fig:model_edit_result} shows that scaling a token’s IDF upward or downward produces a monotonic change in the relevance score.  
 
Taken together with Section~\ref{sec:circuit-function-approximatio} (which shows that Relevance Scoring Heads perform BM25-style computations on natural IR datasets), these results provide stronger causal evidence for the Semantic Scoring hypothesis.

\subsection{Independence between IDF and Soft-TF}
In this subsection, we check whether term-frequency (TF) and inverse document frequency (IDF) factors are fully separable. 

\paragraph{Low correlation between IDF Soft-TF.} 
If soft-TF scores are already IDF-weighted, we would expect their one-dimensional values to correlate at least mildly with $U0$. However, across all 13 Matching Heads, the average Pearson correlation between $U0$ and each head’s Matching-Score is low ($r = 0.143$, $p < 0.01$), suggesting minimal dependence.  

\paragraph{Relevance Scoring Heads Do Not Inherit IDF Information From Matching Heads.} 
Path patching from the Matching Heads to the query vectors of the Relevance Scoring Heads produces almost no change in output score. This indicates that the soft-TF signals produced by the Matching Heads do not strongly convey IDF information, leaving the Relevance Scoring Heads without sufficient signal to prioritize which tokens should receive higher weight.

\section{Generalization Across IR Datasets}
\label{sec:linear_generalization}
We previously test our linear model on an MSMARCO-based dataset with a fixed number of query tokens as an initial proof of concept. Now, we ask the question: how well does this linear model represent the cross-encoder's relevance computation on datasets it was not trained on? In this section, we extend our analysis to 12 different IR datasets\footnote{Due to computational resources, we use a subset of BEIR datasets~\cite {beir} (ArguAna, Climate-FEVER, FEVER, FiQA-2018, HotpotQA, NFCorpus, NQ, Quora, SCIDOCS, SciFact, TREC-COVID, Touche-2020), but the chosen ones sufficiently cover a wide variety of domains.} and investigate various query lengths.

\textbf{Dataset.}
Since our linear model relies on fixed features based on query length, we stratify the dataset based on query length.  To ensure sufficient samples, we aggregate samples from 12 datasets and only retain groups with more than 500 samples. We incorporate groups with query lengths from 1 to 22, which corresponds to the 99.95th percentile of MSMARCO's query length distribution to exclude outliers with excessively long queries that the model has rarely seen during training. This process results in a total of 19 groups, comprising 278194 samples, with an average group size of 14641.790 samples. While this approach may be unconventional for retrieval, our goal is to simply demonstrate consistency between the linear model's relevance score and the cross-encoder to confirm the hypothesized function of the Relevance Scoring Heads.

First, as cross-encoders are typically used for re-ranking tasks, we follow the classic re-ranking setup by first retrieving a candidate set of documents (top 10) with BM25 over all queries across the test collection of the 12 datasets. These candidate sets are then split into train-test groups using an 80/20 ratio.

\textbf{Experiments and Results.}
We train 22 linear regression models, one for each query length (1-22), to predict the cross-encoder's relevance scores. We evaluate each model using: (1) Pearson Correlation: quantifies the correlation between predicted and actual relevance scores; (2) Spearman Rank Correlation: assesses the consistency of the ranked lists between predictions and cross-encoder outputs; (3) NDCG@10: measures ranking effectiveness and ensures no significant effectiveness discrepancies. 
We include BM25 and a randomized set of linear regression features as baselines.

Table \ref{tab:table_validation} shows the results, and we report median values to account for observable skewness caused by outliers in the data. We observe a Pearson correlation of ranking scores with a median of 0.8401 (median \textit{p} < 0.001), a Spearman rank correlation of 0.7619 (median \textit{p} = 0.072), and an 88.4\% alignment with cross-encoder effectiveness in terms of NDCG@10. 

The experimental results confirm the hypothesized function of the Relevance Scoring Heads (\S\ref{sec:relevance_scoring_heads}). Since this linear model summarizes the whole circuit as it is structured to incorporate both the computation and the necessary components, the high correlation with the cross-encoder's effectiveness shows that our circuit understanding has captured the core part of the cross-encoder's relevance ranking mechanism.

\section{Additional Discussion}

\textbf{Towards a holistic approach in axiomatic analysis.}
Previous research investigating neural IR models' adherence to IR axioms shows that BERT does not adhere to most of them, leading to concerns about the limitations of current axiom definitions and the need for new ones \cite{camara2020diagnosing}. Our findings offer an explanation for these shortcomings, suggesting that the challenge in axiomatic analysis lies with the way it has been applied, rather than the axioms themselves. BERT's ``rediscovery'' of BM25 indicates that it leverages a combination of retrieval heuristics to compute relevance, whereas previous analysis treats axioms independently. Going forward, we may consider analyzing axioms in combinations to better reflect the multidimensional nature of relevance. 

\clearpage
\begin{figure*}
    \centering
    \includegraphics[width=1\linewidth]{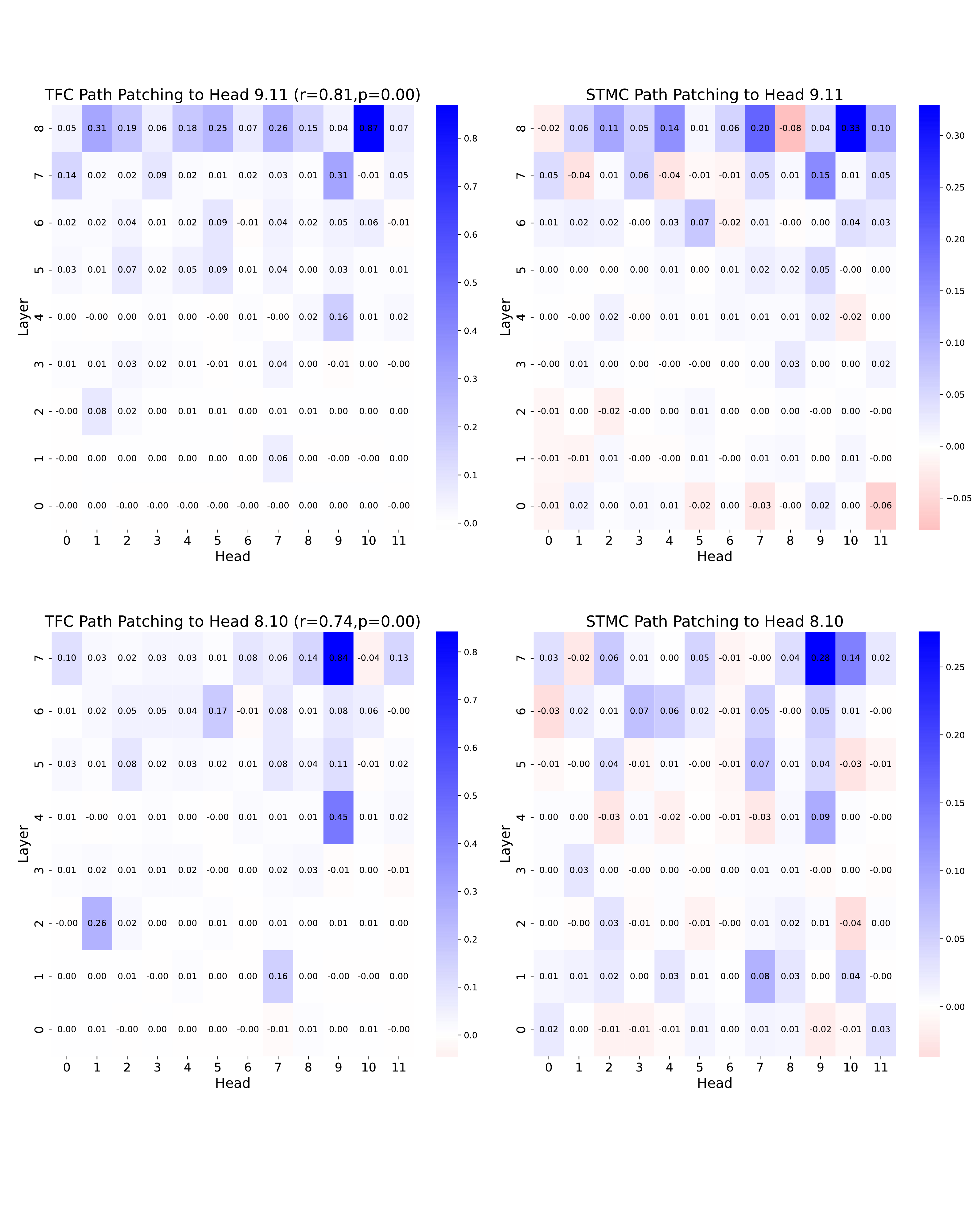}
    \caption{Path Patching results from upstream attention heads to Query Contextualization heads. }
    \label{fig:appendix_path_patching_to_qch}
\end{figure*}


\end{document}